\documentclass[sigconf, nonacm]{acmart}

\usepackage{graphicx} 
\usepackage{amsmath}  
\usepackage{algorithm} 
\usepackage{algpseudocode}
\usepackage{amsthm}
\usepackage{bbm}
\usepackage{amsfonts}
\usepackage{xcolor}
\usepackage{setspace}
\usepackage{multirow}
\usepackage{booktabs}
\usepackage{multirow}    
\usepackage{booktabs}    
\usepackage{tabularx}    
\usepackage{adjustbox}   
\usepackage{xcolor}      
\definecolor{olivegreen}{rgb}{0, 0.6, 0}
\usepackage{caption}     
\usepackage{makecell}
\usepackage{caption}
\usepackage{subcaption} 
\usepackage{float}      
\usepackage{tcolorbox}
\usepackage{enumitem}
\usepackage{mathrsfs} 
\usepackage{tikz}
\usepackage[capitalize]{cleveref}
\usepackage{dblfloatfix}
\newtheorem{definition}{Definition}
\newtheorem{theorem}{Theorem}
\usepackage{xspace}
\usepackage{siunitx}

\newcommand{\thiswork}{AGIS\xspace}
\newcommand{\SAMPLING}{Structure-Informed Neighbor Sampling\xspace}
\newcommand{\Sampling}{Structure-informed neighbor sampling\xspace}
\newcommand{\sampling}{structure-informed neighbor sampling\xspace}

\newcommand{\fapx}{$\mathbf{f}_{\text{approx}}$\xspace}
\newcommand{\funi}{$\mathbf{f}_{\text{uniform}}$\xspace}
\newcommand{\fideal}{$\mathbf{f}_{\text{ideal}}$\xspace}

\crefname{figure}{Figure}{Figures}
\crefname{algorithm}{Algorithm}{Algorithms}
\crefname{equation}{Eq.}{Eqs.}
\crefname{table}{Table}{Tables}

\definecolor{olivegreen}{rgb}{0, 0.6, 0}
\definecolor{redorange}{HTML}{FF5349}
\definecolor{blue(ncs)}{rgb}{0.0, 0.53, 0.74}
\definecolor{navy}{HTML}{273BE2}
\definecolor{greenyellow}{HTML}{DDE576}
\usepackage{xcolor}

\usepackage{hyperref}
\usepackage{marginnote}

\newcommand\vldbdoi{10.14778/3773749.3773761}
\newcommand\vldbpages{238 - 251}
\newcommand\vldbvolume{19}
\newcommand\vldbissue{2}
\newcommand\vldbyear{2025}
\newcommand\vldbauthors{\authors}
\newcommand\vldbtitle{\shorttitle} 
\newcommand\vldbavailabilityurl{https://github.com/syleeKR/AGIS}
\newcommand\vldbpagestyle{empty} 

\begin{document}

\title{\thiswork: Fast Approximate Graph Pattern Mining with Structure-Informed Sampling}
\author{Seoyong Lee}
 \affiliation{%
   \institution{Seoul National University}
   \city{Seoul}
   \country{South Korea}}
 \email{sylee2685@snu.ac.kr}

 \author{Jinho Lee}
 \affiliation{%
   \institution{Seoul National University}
   \city{Seoul}
   \country{South Korea}}
\email{leejinho@snu.ac.kr}

\begin{abstract}
Approximate Graph Pattern Mining (AGPM) is essential for analyzing large-scale graphs where exact counting is computationally prohibitive. While there exist numerous sampling-based AGPM systems, they all rely on uniform sampling and overlook the underlying probability distribution. This limitation restricts their scalability to a broader range of patterns.

In this paper, we introduce AGIS, an extremely fast AGPM system capable of counting arbitrary patterns from huge graphs. AGIS employs \sampling, a novel sampling technique that deviates from uniformness but allocates specific sampling probabilities based on the pattern structure. We first derive the ideal sampling distribution for AGPM and then present a practical method to approximate it. Furthermore, we develop a method that balances convergence speed and computational overhead, determining when to use the approximated distribution. 

Experimental results demonstrate that AGIS significantly outperforms the state-of-the-art AGPM system, achieving 
28.5
$\times$ geometric mean speedup and more than 100,000$\times$ speedup in specific cases. Furthermore, AGIS is the only AGPM system that scales to graphs with tens of billions of edges and robustly handles diverse patterns, successfully providing accurate estimates within seconds. We will open-source AGIS to encourage further research in this field.

\end{abstract}

\maketitle

\pagestyle{\vldbpagestyle}
\begingroup\small\noindent\raggedright\textbf{PVLDB Reference Format:}\\
\vldbauthors. \vldbtitle. PVLDB, \vldbvolume(\vldbissue): \vldbpages, \vldbyear.\\
\href{https://doi.org/\vldbdoi}{doi:\vldbdoi}
\endgroup
\begingroup
\renewcommand\thefootnote{}\footnote{\noindent
This work is licensed under the Creative Commons BY-NC-ND 4.0 International License. Visit \url{https://creativecommons.org/licenses/by-nc-nd/4.0/} to view a copy of this license. For any use beyond those covered by this license, obtain permission by emailing \href{mailto:info@vldb.org}{info@vldb.org}. Copyright is held by the owner/author(s). Publication rights licensed to the VLDB Endowment. \\
\raggedright Proceedings of the VLDB Endowment, Vol. \vldbvolume, No. \vldbissue\ %
ISSN 2150-8097. \\
\href{https://doi.org/\vldbdoi}{doi:\vldbdoi} \\
}\addtocounter{footnote}{-1}\endgroup

\ifdefempty{\vldbavailabilityurl}{}{
\vspace{.3cm}
\begingroup\small\noindent\raggedright\textbf{PVLDB Artifact Availability:}\\
The source code, data, and/or other artifacts have been made available at \url{\vldbavailabilityurl}.
\endgroup
}

\section{Introduction}
\label{sec:intro}

Graph pattern mining (GPM)~\cite{GPM, arabesque,rstream,peregrine,g2miner,graphpi,automine,fractal} is one of the most time-consuming applications within graph processing workloads. 
Given a large data graph and a relatively small pattern graph, GPM searches for embeddings of the pattern from the data graph.
Explicit pattern counts provide concise, deterministic summaries of higher‑order graph structure,
serving as fundamental descriptive statistics that underpin rigorous statistical analysis~\cite{ergm}.

Moreover, as graphs are ubiquitous, pattern mining is employed across diverse domains, including bioinformatics~\cite{biology1,biology2,biology3}, social network analysis~\cite{social_network_community_identification,social_network_structures}, and chemical compound classification~\cite{chemical_compound_classification}.
More importantly, pattern counts are essential for interpretability and accountability in high‑stakes settings, such as fraud detection and cybersecurity, where auditability and regulatory compliance depend on precise, reproducible evidence~\cite{anomaly_detection,cyber_security,RUSH24,MOTIF24}.
Despite its significance in numerous real world scenarios, GPM suffers from its inherent high computational complexity, making it hard to scale to large graphs or complex patterns.  
This challenge is especially pertinent in the current era, as modern applications demand low-latency analysis on massive graphs ~\cite{RUSH24,MOTIF24,graphs_are_everywhere}.

One popular approach to alleviate this is approximate graph pattern mining (AGPM).
Using the fact that GPM tasks, such as frequent subgraph mining~\cite{fsm_gspan} and motif counting~\cite{motif}, are employed in applications where estimated embedding counts suffice, AGPM computes approximate embedding counts for a given pattern within specified error bounds and confidence intervals.
While there exist several seminal AGPM systems ~\cite{arya, asap, scalegpm}, they still have difficulties as the data graphs and patterns become larger.
Upon characterizing them, we identify that certain patterns experience especially larger slowdowns. 
Our analysis shows that this can be attributed to the scale-free distribution~\cite{scalefree, scalefree_empirical, scalefree2} often found in real-world graphs. 
Such a distribution leads to a large variance in the embedding count depending on which part of the data graph is sampled. 
Because it makes the estimated count difficult to converge, this results in a long execution time for AGPM systems. 

Fortunately, we identify that such a problem can be alleviated by assigning different probabilities to each sampling candidate. 
While existing methods also try to reduce the estimation variances, they often focus on narrowing down the candidates by only considering neighbors of already sampled vertices~\cite{asap}, or neighbors that satisfy certain constraints~\cite{scalegpm}.
However, once the candidates set is constructed, the probability of each candidate to be sampled is uniform.

Instead of applying naive uniform sampling probability distribution, we propose to use uneven distribution where the probability to sample each candidate vertex is proportional to the number of potential embeddings found by choosing the vertex.
In theory, this leads to a correct count without variance with a single sampler.
However, this requires prior knowledge about the true embedding count, which defeats the purpose of AGPM.
To construct an appropriate sampling probability distribution without relying on the true embedding count, we develop a method to approximate the distribution based on the structural characteristics of the data and pattern graph.
By carefully examining the nearby structures of the target vertex to be sampled, we calculate a distribution close to the ideal one with a small overhead.

Based on this, we propose \thiswork, a fast and scalable AGPM system.
At the heart of \thiswork is \emph{\sampling}.
\Sampling leverages the calculated distribution and strategically determines when to apply it, thereby achieving an optimal balance between faster convergence and minimal computational overhead.
In addition, we provide a heuristic to construct a matching order such that the benefit of \sampling is maximized.
We evaluate \thiswork over a various data graph and patterns, against several state-of-the-art baselines.
Experimental results show that \thiswork outperforms all the baselines, sometimes by an order of several magnitudes. 
\thiswork sets a new state-of-the-art, further extending the applicability of AGPM methodology.

Our contributions can be summarized as follows:
\begin{itemize}[]
    \item We show that by assigning different probabilities to each vertex in the sampling candidates, significantly faster convergence can be obtained for AGPM.
    \item We develop a method for constructing an approximate ideal sampling distribution and a heuristic decision process that determines when to apply it.
    \item We build \thiswork, a fast AGPM system that significantly outperforms prior art over a diverse set of experiments.
\end{itemize}

\section{Background}

In this section, we provide a comprehensive background on approximate graph mining and its core strategies in existing work.
We use \( G = (V_G, E_G) \) and \( P = (V_P, E_P) \) to represent the data and pattern graphs, respectively.  We assume that both graphs are undirected; thus, \( E_G \) and \( E_P \) are sets of unordered pairs \( \{a, b\} \). For a vertex \( v \), \( d(v) \) denotes its degree, \( \mathcal{N}(v) \) its set of neighbor vertices, and \( v.\text{id} \) its id. We use \( v \) for vertices in \( G \) and \( u \) for vertices in \( P \).

\subsection{Approximate Graph Mining Systems}
\label{sec:back:basic}

\textbf{Graph Pattern Mining}. Graph Pattern Mining (GPM) is the problem of finding all \textit{embeddings}, i.e., matches, of a pattern \( P \) within a given graph \( G \). Formally, we define an embedding as a subgraph isomorphism from \( P \) to \( G \). Specifically, it is a one-to-one mapping \( \mathcal{M} : V_P \rightarrow V_G \) such that if \( (u_i, u_j) \in E_P \), then \( (\mathcal{M}(u_i), \mathcal{M}(u_j)) \in E_G \). We let \( C(G,P) \) denote the total number of such embeddings. Example tasks in GPM include subgraph counting~\cite{fractal}, subgraph listing 
(subgraph matching)
~\cite{fractal}, motif counting~\cite{motif}, and frequent subgraph mining~\cite{fsm_gspan}.

\begin{figure}
\centering
\includegraphics[width=.9\columnwidth]{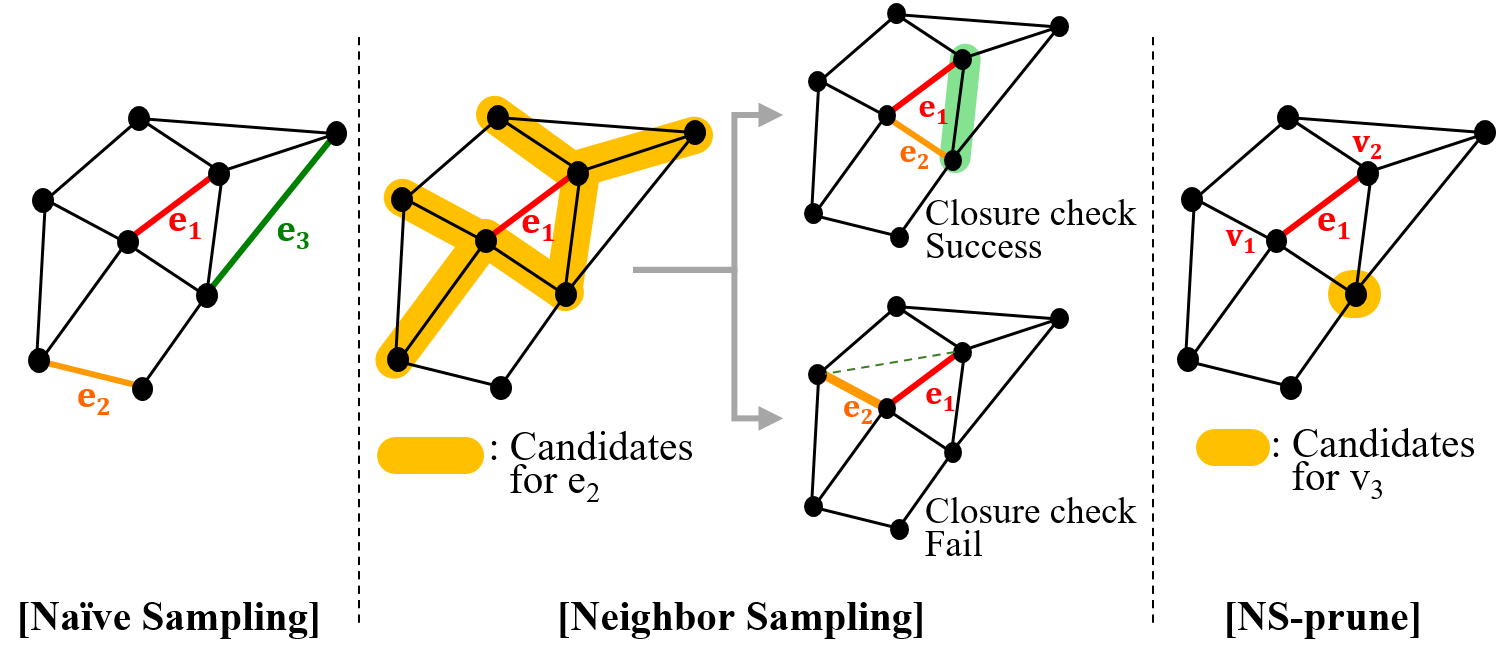}
\caption{Sampling strategies of existing AGPM systems.}
\label{fig:background}
\end{figure}

\noindent \textbf{Approximate Graph Pattern Mining}. Approximate Graph Pattern Mining (AGPM) addresses the subgraph counting problem in an approximate manner. Given a graph \( G \), a pattern \( P \), an error bound \( \epsilon \), and a confidence level \( 1 - \delta \), AGPM seeks to return a \( (1 \pm \epsilon) \) approximation of the exact number of embeddings $C(G,P)$, with probability at least \( 1 - \delta \). AGPM systems enable users to mine arbitrary patterns using two major components: (1) a general sampling method applicable to any pattern, and (2) a convergence detection method that determines when to terminate sampling given 
\( (\epsilon, \delta) \).

\subsection{Sampling Strategies of AGPM}
\label{sec:back:sampling}

To obtain approximate values for $C(G,P)$, existing approaches~\cite{asap, arya, scalegpm} employ the sampling method.
As a simple example, suppose we were to approximate the number of triangles in graph $G$. 
In a straightforward \texttt{Naive} sampling method shown in \cref{fig:background}, a sampler randomly samples three edges \( e_1, e_2, e_3 \) and checks if they form a triangle. If they do form a triangle, the sampler estimates the number of triangles as \( |E_G|^3 \) since each of the three edges is sampled randomly with a probability of \( \frac{1}{|E_G|} \). 
Otherwise, the sampler estimates the number of triangles as 0. 
The average from many such samplers will converge toward the true triangle count.

One significant issue with the \texttt{Naive} sampling method is its large variance between the sampler outputs. 
Because each sampler outputs either $0$ or the substantially large value $|E_G|^3$, the average of samplers converges slowly.
Neighbor Sampling (\texttt{NS})~\cite{ns} addresses this problem by leveraging the connectivity of the pattern graph. 
Specifically, a \texttt{NS} sampler operates as follows: 
it first samples an edge \( e_1 \) uniformly at random from \( E_G \), similar to the sampler using \texttt{Naive} method. However, 
instead of sampling the next edge uniformly from \( E_G \), it samples \( e_2 \) from the set of edges adjacent to \( e_1 \). Then, it performs a \emph{closure check}, which determines whether an edge \( e_3 \) exists between the non-shared vertices of \( e_1 \) and \( e_2 \). This procedure confirms whether a triangle embedding can be formed from the sampled edges. The probability is \( \frac{1}{|E_G| \cdot |N(e_1)|} \), where \( N(e_1) \) denotes the set of neighboring edges of \( e_1 \).
Because \( |E_G| \cdot |N(e_1)| \) is smaller and closer to the true triangle count compared to \( |E_G|^3 \), the  \texttt{NS} sampler outputs have lower variance and converges faster.


The aforementioned methods can easily be generalized to an arbitrary pattern $P$.
A sampler sequentially samples vertices until the number of sampled vertices reaches \( |V_P| \). 
Then, the probability $p$ of this sampling sequence is given by 
\begin{equation*}
 p = \prod_{i=1}^{|V_P|} Pr(v_i | v_1, \dots , v_{i-1}), \label{eq:general_nsprune}
\end{equation*}
where $Pr(v_i | v_1, \dots , v_{i-1})$ denotes the probability to sample $v_i$ given $v_1, \dots v_{i-1}$ is sampled. 
Each sampler returns \( \frac{1}{p} \) if the vertices form an embedding of \( P \) and returns 0 otherwise. 
This sampler serves as an unbiased estimator for \( C(G,P) \)~\cite{asap}. 

The \texttt{NS} method can further be modified by imposing restrictions on neighbor selection~\cite{4vertexcounting, motifbeyond5nodes}.
For example, when sampling the \texttt{k-star} pattern (a pattern where $k$ vertices are connected to a central vertex), we can sample the next edge only from the neighbors of the central vertex to maximize the success rate of the closure check.
Building on this idea, ScaleGPM~\cite{scalegpm} proposes a modified version of \texttt{NS} called \texttt{NS-prune}. This method not only exploits the connectivity of the pattern \( P \) but also leverages its specific topology. It introduces two concepts of exact GPM systems:

\textbf{Matching Order}. A matching order~\cite{pbe, g2miner} \( \pi: \{1, \dots, |V_P|\} \to V_P \) is a permutation of the pattern vertices that specifies the order in which the vertices of \( P \) are matched to the graph \( G \). We denote \( u_i = \pi(i) \in V_P\) as the \( i \)-th vertex of \( P \) in the matching order. Correspondingly, we write \( v_i \in V_G \) as the \( i \)-th sampled vertex of \( G \). 



\textbf{Restriction Set}. Given a matching order \( \pi \), a restriction set~\cite{graphzero} \( \mathcal{R}^\pi \) is a set of ordered pairs of pattern vertices \( (u_i, u_j) \). 
$\mathcal{R}^\pi$ is applied with \textit{symmetry breaking}~\cite{graphzero}, which enforces the condition \( v_i.\text{id} < v_j.\text{id} \) for all \( (u_i, u_j) \in \mathcal{R}^\pi \).
This is beneficial for exact mining systems, as it reduces the search space by exactly the number of automorphisms of $P$. Although this approach does not enumerate all embeddings, the total count can be recovered by simply multiplying the result accordingly.

\texttt{NS-prune} utilizes these concepts as follows. Taking the triangle pattern again as an example, it first samples \( e_1 = \{v_1, v_2\} \) randomly like the prior methods. 
Then, instead of sampling any \( e_2 \) from the neighbors, it leverages the fact that \( u_3 \) is connected to \( u_1 \) and \( u_2 \). 
Thus, the next vertex \( v_3 \) must be selected from the intersection of the neighbors of \( v_1 \) and \( v_2 \), that is, \( v_3 \in \mathcal{N}(v_1) \cap \mathcal{N}(v_2) \).  
By enforcing these \emph{connectivity constraints} at each step, the success rate of finding an embedding increases, leading to faster convergence. Additionally, the triangle pattern has 3! number of automorphisms, having a restriction set $\mathcal{R}^\pi = \{(u_2,u_1), (u_3,u_2), (u_3,u_1)\}$. By applying symmetry breaking, the condition \( v_3.\text{id} < \min(v_1.\text{id}, v_2.\text{id}) \) is enforced. The enforcement of these restrictions reduces the set size of possible vertices. Since the sampling probability is equal to the inverse of the set size, 
the estimator's variance is reduced.
Lastly, for general patterns, the \texttt{NS-prune} avoids sampling vertices from the previously sampled vertices. 

Considering the above techniques, \texttt{NS-prune} samples the last vertex \( v_3 \) uniformly from the candidate set
\begin{equation}
S = \left\{ v \mid v \in \mathcal{N}(v_1) \cap \mathcal{N}(v_2),\ v.id < \min(v_1.id,\ v_2.id) \right\}.\notag
\end{equation}



It then returns \( |E_G| \cdot |S| \) as the sampler's output. In 
ScaleGPM, this value serves as an unbiased estimate of \( C(G, P) \) divided by the number of automorphisms of the pattern, due to the use of the restriction set. Since the sampler produces non-zero outputs more frequently, its variance is reduced, leading to faster convergence.


    
    
    
    
    

\subsection{Convergence Method}
\label{sec:back:convergence}
An important issue to be addressed from AGPM is to ensure convergence of $C(G,P)$ within the error bound $\epsilon$.
ASAP~\cite{asap} and Arya~\cite{arya} utilize the error-latency profile (ELP) heuristic that predetermines the number of samplers needed for convergence based on concentration inequalities. 
For example, Arya uses 
Chebyshev’s inequality~\cite{aryaelp} to obtain the number of samplers as follows:
\begin{equation*}
\text{\# samplers needed} = \frac{K \times |E_G|^\rho}{C(G,P) \times \epsilon^2 \times \delta},
\end{equation*}
where $\rho$ represents a pattern specific value, and \( K \) is a constant:
However, the equation requires the value of $C(G,P)$, where the constant $K$ is also unknown. 
To address this, they rely on sampling from a smaller subgraph of $G$ to estimate $K$ and $C(G,P)$.

\begin{algorithm}
\caption{Online Convergence Detection}\label{alg:convergences}
\begin{spacing}{1.3}
\begin{algorithmic}[1]
\Procedure{Converged}{$\epsilon$, $\delta$, $\mathbf{L}$}
    \State $N \gets |\mathbf{L}|$ \Comment{\( \mathbf{L} \) is the list of sampler outputs}
    \State $\mu \gets \frac{1}{N} \sum_{i=1}^N L_i$, \quad $\sigma^2 \gets \frac{1}{N} \sum_{i=1}^N L_i^2 - \mu^2$ 
    \State $\hat{\epsilon} \gets  \Phi^{-1}\big(1 - \frac{\delta}{2}\big)  \cdot \frac{\sigma}{\sqrt{N}} \cdot \frac{1}{\mu}$  \Comment{Estimated error}
    \State \Return $\hat{\epsilon} \leq \epsilon$  
\EndProcedure
\end{algorithmic}
\end{spacing}
\end{algorithm}

Unfortunately, the method based on ELP lacks a theoretical guarantee as it relies on a subgraph whose $K$ and $C(G,P)$ will be different from that of the original $G$. To address this problem, ScaleGPM~\cite{scalegpm} proposes a mathematically sound method to determine the termination of the sampling, as described in Algorithm~\ref{alg:convergences} where $\Phi$ is the CDF of the standard normal distribution. Instead of predetermining the number of samplers, it detects convergence by using the mean and variance of the sampler results. The method is proven for arbitrary sampling-based techniques that provide unbiased estimates.



\section{Motivation}
\subsection{Limitations of Existing Methods}\label{sec:limitations}

%
%
%

\begin{figure}[t]
    \centering
    \includegraphics[width=0.49\textwidth]{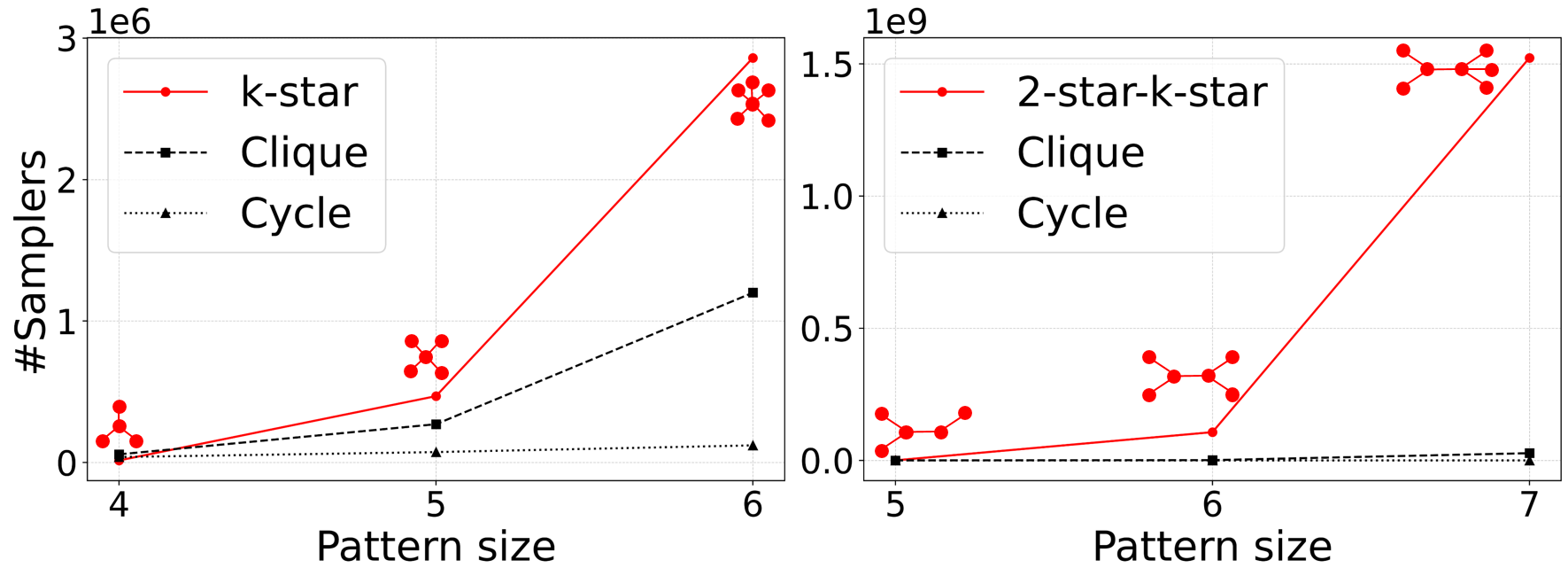} 
    \caption{Number of samplers needed for convergence. LiveJournal graph, using $\epsilon = 0.1$, $\delta = 0.01$.}
    \label{fig:motivation}
\end{figure}

Even though AGPM systems achieve significantly shorter execution times compared to exact GPM systems, they are known to suffer from low convergence speed as the pattern size increases.
Specifically, we find that the severity of such a problem is exacerbated by certain types of patterns. One example is patterns with many bridges~\cite{bridge}, which are defined as edges whose removal would result in the separation of the graph.
\cref{fig:motivation} illustrates the number of samplers required for convergence on two patterns---\texttt{k-star} and \texttt{2-star-k-star}---on Livejournal~\cite{lj} graph using ScaleGPM.
Compared to the clique and cycle patterns with the same number of vertices, the two patterns we examine require more samplers to converge, with much steeper growth rates. 
Considering that a clique has far more edges than the examined patterns at an equal number of vertices, the results indicate that there exists some inefficiencies on top of pure complexity growth.

One reason for this is the scale-free distribution \cite{scalefree,scalefree2, scalefree_empirical} observed in many real-world graphs, where the degree distribution of vertices is highly skewed. 
Suppose we are sampling the \(i\)-th vertex \(v_i\), following a bridge \(\{u_j, u_i\} \in E_p\) (with \(j < i\)). 
Accordingly, the connectivity constraints from \cref{sec:back:sampling} cannot be applied because, by the definition of a bridge, no other vertex \(u\) with \(\pi^{-1}(u) < i\), except \(u_j\), shares an edge with \(u_i\).
Therefore, omitting symmetry breaking for simplicity, $|S|$ is approximately equal to $d(v_j)$, the degree of \(v_j\).
%
Since the degrees vary significantly in graphs with a scale-free distribution, the sampler's outputs will also have a high variance in the estimate and ultimately slower convergence. Moreover, if vertex $u_j$ in the pattern is connected to $k$ bridges, the sampler's output will be roughly proportional to \(d(v_j)^{k-1}\) (see \cref{sec:moti:prob}), which exacerbates this issue.


\subsection{Rethinking Sampling Probability}
\label{sec:moti:prob}

In the three sampling methods described in \cref{sec:back:sampling} (\texttt{Naive}, \texttt{NS}, and \texttt{NS-prune}) improvements were made by reducing the size of the sampling set. 
With a more promising candidate set, the probability of finding an embedding increases, leading to faster convergence.

These three methods differ in how they form a candidate set $S$, but they all assign uniform sampling probability to elements within $S$. 
In the \texttt{Naive} sampling method, the candidate set is the entire edge set \( E_G \) with the uniform sampling probability.
In \texttt{NS}, the candidate set is narrowed to the edges neighboring the already-sampled vertices. 
For \texttt{NS-prune}, the candidate set is further refined only to the vertices that satisfy the connectivity constraints and symmetry-breaking restrictions.
%
Assigning uniform probabilities to these set elements is a reasonable choice.
However, such a choice is not mandatory. 
%
In fact, these methods can already be interpreted as assigning non-uniform probabilities restricted to the first vertex.
All three methods uniformly sample a random edge at the beginning which can be interpreted as having all vertices in the set and having probability proportional to their degrees.
%

From these observations, we propose to further add an extra step to calculate a tailored sampling distribution, allowing us to assign distinct probabilities to each element in a way that it achieves faster convergence.
As a motivational example, consider the \texttt{k-star} pattern. Suppose we use \texttt{NS-prune} without symmetry breaking and start with the central vertex. Since sampling the first edge uniformly is equivalent to sampling the first vertex with a probability proportional to its degree, a central vertex $v$ is sampled with probability \( \frac{d(v)}{2|E_G|} \). 
Subsequently, for the remaining $k$ outer vertices, the candidate set is defined as the set of all neighbors of the central vertex, excluding any vertices that have already been sampled. Thus the sampler outputs
\(
\frac{2|E_G|}{d(v)} \times d(v) \times \bigl( d(v) - 1 \bigr) \times \dots \times \bigl( d(v) - (k - 1) \bigr).
\)
Since \( d(v) \) varies greatly per vertex, it would need numerous iterations to achieve convergence. 

On the other hand, an interesting aspect of \texttt{k-star} is that a vertex $v$ can form exactly $\binom{d(v)}{k}$ such patterns.
What if we discard uniform sampling and instead sample the first vertex $v$ with probability proportional to \( \binom{d(v)}{k} \); that is, \( p = \frac{\binom{d(v)}{k}}{\sum_{x \in V_G}\binom{d(x)}{k}} \) ? The sampler's output will then be
\(
\frac{\sum_{x \in V_G}\binom{d(x)}{k}}{\binom{d(v)}{k}} \times d(v) \times \bigl( d(v) - 1 \bigr) \times \dots \times \bigl( d(v) - (k - 1) \bigr) = {\sum_{x \in V_G}\binom{d(x)}{k}} \times k!,
\)
which is a constant value equal to $C(G,P)$. This means that our sampler will always return the same value, resulting in zero variance, and convergence to $C(G,P)$ will occur immediately with a \emph{single} sampler.
%

\section{Building Sampling Distributions}
While the result from \cref{sec:moti:prob} is highly appealing, two key challenges arise when applying the principle to general patterns. First, while identifying the ideal distribution for a simple pattern like \texttt{k-star} is straightforward---we can precompute the sampler outputs exactly---it is unclear how to find such a distribution for a general pattern. Second, even if we could determine the ideal distribution for a complex pattern, computing it directly could be prohibitively expensive, diminishing the advantages of faster convergence.
In this section, 
we first discuss an ideal distribution for a general pattern (\cref{sec:method:ideal}), how to approximate it (\cref{sec:method:approx}), 
the theoretical intuition behind it (\cref{sec:method:sample_complexity})
and how to preprocess the data graphs accordingly (\cref{sec:method:preprocessing}).

\newcommand{\concat}{\circ}


\subsection{Ideal Distribution for General Patterns}
\label{sec:method:ideal}
In this section, we first provide the ideal sampling probability distribution for a generalized pattern. 

\begin{definition}[Sampling Trajectory]
Given a graph \( G \), a \emph{sampling trajectory} \( \tau \) is an ordered sequence of vertices sampled from \( V_G \), denoted by
\[
\tau = ( v_1, v_2, \dots, v_k ),
\]
where \( v_i \in V_G \) is the \( i \)-th sampled vertex.

If we sample a vertex \( v \in V_G \) after \( \tau \), we denote the new sampling trajectory as \( \tau' = \tau \concat v \), where \( \concat \) represents concatenation.
\end{definition}

\begin{definition}[Successful Trajectory]

A sampling trajectory \( \tau \) is \emph{successful} if it represents an embedding. That is,
\[
|\tau| = |V_P| \quad \text{and} \quad \forall\, (u_i, u_j) \in E_P,\ (v_i, v_j) \in E_G.
\]
\end{definition}

\begin{definition}[Number of Successful Extensions]
Given a sampling trajectory \( \tau = ( v_1, v_2, \dots, v_k ) \) with \( 0 \leq |\tau| \leq |V_P| \), the \emph{number of successful extensions} \( n_\tau \) is the total number of distinct successful trajectories that extend \( \tau \) by adding vertices from \( V_G \). Formally,
\[
n_\tau = \left| \left\{ \tau' \in {V_G}^{|V_P|}\ \bigg|\ \begin{aligned}
& \tau' = ( v_1, v_2, \dots, v_k, v_{k+1}, \dots, v_{|V_P|} )\\
& \tau' \text{ is a successful trajectory}
\end{aligned} \right\} \right|.
\]
\end{definition}
For example, if $\tau = \emptyset$, the value $n_{\tau}$ is $C(G,P)$. Also for  $\tau$ with $|\tau| = |V_P|$, $n_\tau$ is either 1 ($\tau$ is successful) or 0 (is not).

\begin{theorem}[Ideal Sampling Distribution]\label{thm:ideal_sampling_distribution}
Suppose there exists at least one embedding of P in G. Assign a probability to each vertex \( v \in V_G \) such that the probability is proportional to the number of successful extensions that can be generated when \( v \) is selected, $i.e., n_{\tau \concat v}$. Formally,
\[
\mathbf{f}_{\text{ideal}}(v \mid \tau)= \frac{n_{\tau \concat v}}{\sum_{x \in V_G} n_{\tau \concat x}}, \quad \mathbf{f}_{\text{ideal}}(\cdot \mid \tau) \in \mathbb{R}^{|V_G|}.
\]
{Then, a sampler using this probability assignment is ideal in that it returns $C(G,P)$ with zero variance.}
\end{theorem}

\begin{proof}

Suppose we follow the above distribution $\mathbf{f}_{\text{ideal}}$ for sampling. Since $\tau$ is an empty sequence $()$ at the beginning, we sample the first vertex $v_1$ from distribution $\mathbf{f}_{1}$, which is
\[
\mathbf{f}_{1}(v) = \mathbf{f}_{\text{ideal}}(v \mid \tau= ()) = 
\frac{n_{(v)}}{\sum_{x\in V_G} n_{(x)}}.
\]
We sample the first vertex $v_1$ with probability $\frac{n_{(v_1)}}{\sum_{x\in V_G} n_{(x)}}$. Then the next vertex is sampled from 
\[
\mathbf{f}_{2}(v) = \mathbf{f}_{\text{ideal}}(v \mid \tau =(v_1)) =  
\frac{n_{(v_1) \concat v}}{\sum_{x \in V_G} n_{(v_1) \concat x}}.
\]
Repeating this process, at step \( i \), we have the sampling trajectory \( \tau = (v_1, v_2, \dots, v_{i-1}) \), and we sample the \( i \)-th vertex \( v_i \) from the distribution \( \mathbf{f}_{i} \), which is:
\[
\mathbf{f}_{i}(v) = \mathbf{f}_{\text{ideal}}(v \mid \tau = (v_1, \dots, v_{i-1}) ) = 
\frac{n_{\tau \concat v}}{\sum_{x \in V_G} n_{\tau \concat x}}.
\]
Since there exists at least one embedding of \( P \) in \( G \), and we never sample a vertex \( v \) such that \( n_{\tau \concat v} = 0 \), the sampling process continues until \( |\tau| = |V_P| \), and the sampling trajectory \( \tau = (v_1, v_2, \dots, v_{|V_P|}) \) is always a successful one.

Moreover, for any such \( \tau \), the sampler output \( X_\tau \) becomes 
{
\footnotesize
\begin{align}
X_\tau &= \frac{\sum_{x\in V_G} n_{(x)}}{n_{(v_1)}} \cdot 
\frac{\sum_{x\in V_G} n_{(v_1 ) \concat x }}{n_{(v_1, v_2)}} \dots \cdot \frac{\sum_{x\in V_G} n_{(v_1 ,\dots v_{|V_P|-1} ) \concat x} }{n_{ (v_1 ,\dots v_{|V_P|} )}} 
= C(G,P), \notag
\end{align}
}
since $n_\tau=n_{(v_1,\dots,v_{|V_P|})}=1$, $\sum_{x\in V_G} n_{(x)} = C(G,P)$, and the terms cancel out as $\sum_{x\in V_G} n_{ (v_1, \dots\ v_i ) \concat x }  = n_{(v_1, \dots v_i)}$.
The sampler always returns the number of embeddings $C(G,P)$ with variance $0$. 
\end{proof}

\subsection{Approximating the Ideal Distribution}
\label{sec:method:approx}

We have established that sampling from the ideal distribution \(\mathbf{f}_{\text{ideal}}(\cdot \mid \tau)\) yields an unbiased, zero-variance estimate. 
However, constructing the ideal probability distribution requires prior knowledge of the ratios of \( n_{\tau \concat v} \), which necessitates knowing \(C(G,P)\) beforehand. Since this is infeasible, we need an efficient method to build an approximate distribution \(\mathbf{f}_{\text{approx}}(\cdot \mid \tau)\) that closely approximates \(\mathbf{f}_{\text{ideal}}(\cdot \mid \tau)\) and yields unbiased results. 

\subsubsection{Unbiasedness of a sampling distribution}

Prior to constructing \(\mathbf{f}_{\text{approx}}\), a natural question arises: can arbitrary sampling distributions be employed while still preserving the unbiasedness of the resulting estimator? The following theorem affirms this, under specific conditions.

\begin{theorem}\label{thm:unbiased_sampling}
Suppose the sampling distribution \( \mathbf{f}(v \mid \tau) \) assigns a positive probability to every vertex \( v \in V_G \) such that \( n_{\tau \concat v} > 0 \). That is, the distribution ensures that every successful trajectory is reachable. Then the sampler is an unbiased estimator of \( C(G,P) \).
\end{theorem}

\begin{proof}
Let \( X \) be the output of the sampler. By definition, the expectation of \( X \) is given by
\[
\mathbb{E}[X] = \sum_{\tau} \mathbbm{1}_{\{\tau \text{ is successful}\}} X_\tau \, p_\tau.
\]
where \( X_\tau \) represents the sampler's output for trajectory \( \tau \), and \( p_\tau \) denotes the probability that the sampler follows trajectory \( \tau \). This expression holds because a sampler outputs \( X_\tau = 0 \) for any trajectory \( \tau \) that is not successful. 

First, note that each successful trajectory corresponds to a unique embedding of \( P \).
Second, every unique embedding is reachable through the sampling process. 
\sloppy{
Therefore, for all \( C(G,P) \) embeddings of $G$, there exist corresponding successful trajectories \( \tau_1, \dots, \tau_{C(G,P)} \).
} 
Thus we can write
\[
\mathbb{E}[X] = \sum_{\tau} \mathbbm{1}_{\left(\tau \text{ is successful}\right)} X_\tau \, p_\tau = \sum_{i=1}^{C(G,P)} X_{\tau_i} p_{\tau_i}.
\]
Also, for each successful \( \tau_i \), the sampler outputs $X_{\tau_i} = \frac{1}{p_{\tau_i}}$, which is well-defined. 
Therefore, the expected value becomes
\[
\mathbb{E}[X] = \sum_{i=1}^{C(G,P)} X_{\tau_i} \, p_{\tau_i} = \sum_{i=1}^{C(G,P)} \frac{1}{p_{\tau_i}} \, p_{\tau_i} = \sum_{i=1}^{C(G,P)} 1 = C(G,P).
\]
Thus, the sampler is an unbiased estimator of \( C(G,P) \).
\end{proof}

\subsubsection{Generalized Approximation Framework}
\label{sec:generalized_approx_framework}
We begin by filtering out unpromising vertices by applying connectivity constraints and excluding already sampled vertices, following \cite{scalegpm}. 
To sample the $i$th vertex (i.e. $i = |\tau|+1$), we define the candidate set $\mathbf{S}_{\tau}$ of promising vertices as
\[
\mathbf{S}_{\tau} = \left\{ v \;\middle|\; v \notin \tau,\; v \in \bigcap_{u_j \in e \in \mathcal{B}, j< i} \mathcal{N}(v_j) \right\},
\]
where $\mathcal{B}$ is the set of backward edges connecting $u_i$ to already sampled vertices, that is
\[
\mathcal{B} = \{\{u_j, u_i\} \in E_P \mid j < i\}.
\]
We assign $\mathbf{f}_{\text{approx}}(v \mid \tau) = 0$ to all \( v \notin \mathbf{S}_{\tau} \), because it is guaranteed that \( n_{\tau \concat v} = 0 \). 

Our goal is now to approximate the ratios of \( n_{\tau \concat v} \) for vertices \( v \in \mathbf{S}_{\tau} \).
The key idea is to consider the structural components of the pattern \( P \) in relation to the data graph \( G \). 
We begin by grouping the vertices based on their distance from \( u_i \) within the subgraph $P_i = (V_i, E_i)$, which comprises the yet-to-be-sampled vertices of \( P \):
\begin{equation} 
V_i = \{ u_j \in V_P \mid j \geq i \}, \  E_i = \{ \{ u_a, u_b \} \in E_P \mid u_a, u_b \in V_i \}.
\label{eq:Pi}
\end{equation}
We then define the \( k \)-hop vertex group as the set of vertices in \( P_i \) that are at a distance \( k \) from \( u_i \) in \( P_i \).
Using this information, we decompose \( n_{\tau \concat v} \) into a product of terms, each accounting for different types of edges divided by \( k \)-hop information:
\begin{equation}
n_{\tau \concat v} \approx  \prod_{k=1}^{D} T[\mathcal{F}_k](v)T[\mathcal{I}_k](v),
\label{eq:general_approximation}
\end{equation}
where
\begin{itemize}[]
    \item \( T[\mathcal{F}_k](v) \) is the term accounting for the set of \emph{forward edges}, \(\mathcal{F}_k\). Specifically, \(\mathcal{F}_k \subseteq E_i\) consists of all edges that connect vertices at hop \((k-1)\) to vertices at hop \(k\).
    \item \( T[\mathcal{I}_k](v) \) is the term accounting for the set of \emph{internal edges}, \(\mathcal{I}_k\). Specifically, \(\mathcal{I}_k \subseteq E_i\) consists of all edges among vertices at the same hop \(k\).
    \item $D$ is the maximum number of hops from $u_i$ to any other connected vertex in $P_i$.
\end{itemize}

Note that \cref{eq:general_approximation} does not account for all edges of $P$. For example, there may be vertices in \( P_i \) that are not connected to \( u_i \), or edges connecting vertices at hop \( k \) to previously sampled vertices. 
Although it is theoretically possible to consider all those edges, we find that focusing on edges of type $\mathcal{F}_k$ and $\mathcal{I}_k$ is sufficiently effective. Moreover, while calculating all $D$-hop terms can yield more accurate probability approximations, we find that it incurs excessive computational overhead with minimal improvements. We find that using the first three terms of the equation provides an effective approximation:
\[
n_{\tau \concat v} \approx T[\mathcal{F}_1](v) \cdot T[\mathcal{I}_1](v) \cdot T[\mathcal{F}_2](v).
\]
%



This introduces a small degree of approximation error; however, such errors affect only the rate of convergence and do not impact the correctness of the embedding count \( C(G, P) \). 
We will present, in~\cref{sec:approximation}, an equation for each term that ensures no vertex is assigned a zero probability unless it is certain, thereby guaranteeing unbiasedness (\cref{thm:unbiased_sampling}).

\subsubsection{Approximation Details} \label{sec:approximation}

\begin{figure}[t]
    \centering
    \includegraphics[width=0.9\columnwidth]{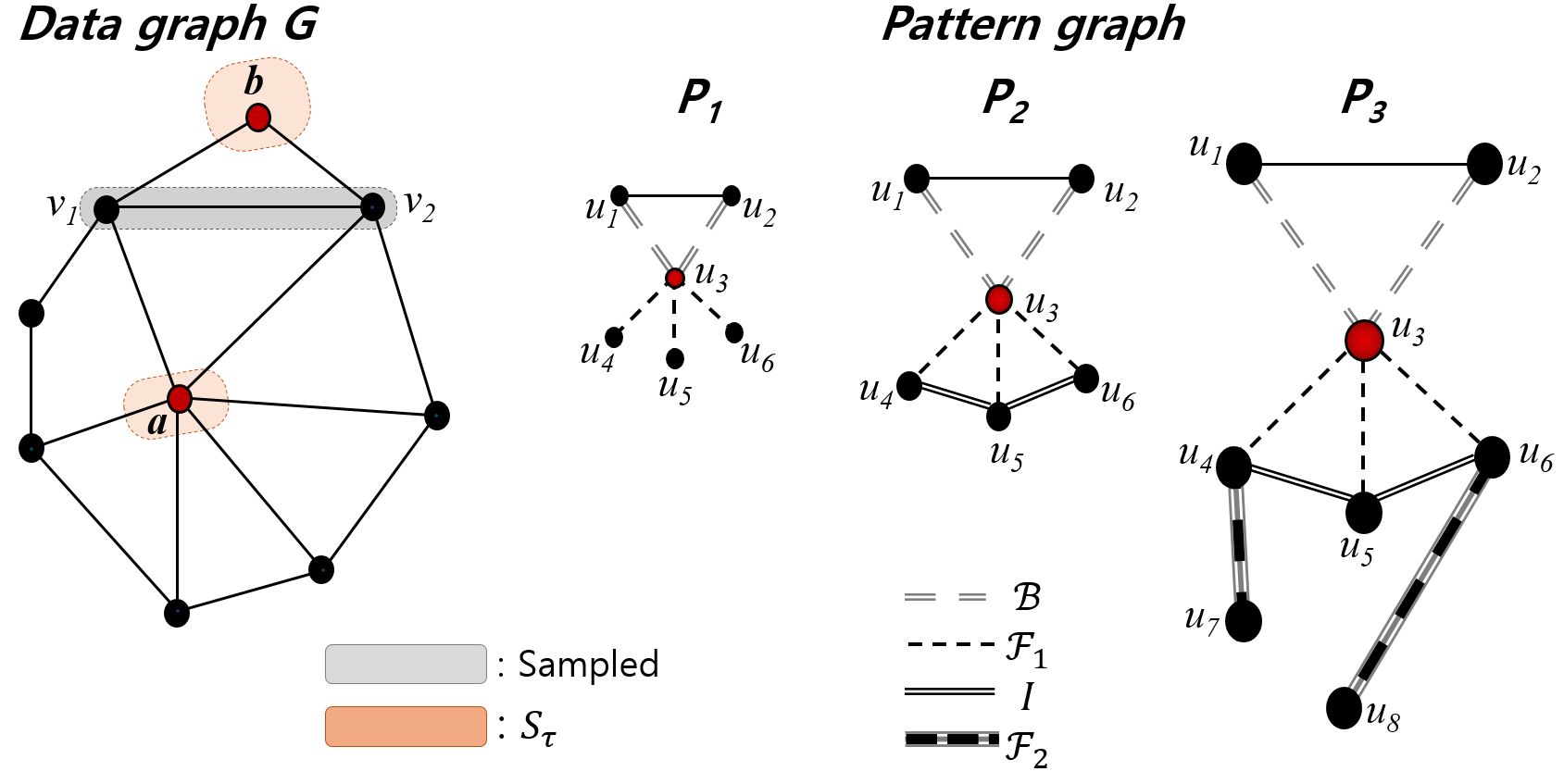} 
    \caption{Example graph and patterns for the approximation.}
    \label{fig:graphexplain}
\end{figure}

In the example illustrated in \cref{fig:graphexplain}, we are sampling the third vertex \( \pi(3) = u_3 \) of the pattern \( P_1,P_2,P_3 \) from the graph \( G \).
Since \( u_3 \) is connected to \( u_1 \) and \( u_2 \), the corresponding vertex \( v_3 \) must be connected to both \( v_1 \) and \( v_2 \) in \( G \). 
Therefore, the candidate set \( \mathbf{S}_{\tau} \) consists of the vertices adjacent to both \( v_1 \) and \( v_2 \) that are not already in \( \tau \); in this case, \( \mathbf{S}_{\tau} = \{ a, b \} \).

Now, consider the term \( T[\mathcal{F}_1](v) \) with pattern \( P_1 \). In $P_1$, we have $\mathcal{B}=\{ \{u_1, u_3\}, \{u_2, u_3\} \} $, and $ \mathcal{F}_1 = \{ \{u_3, u_4\}, \{u_3, u_5\}, \{u_3, u_6\} \} $.
For a vertex \( v \in \mathbf{S}_{\tau} \), it is already connected to the previously sampled vertices with \( |\mathcal{B}| \) edges. Thus, we need to choose \( |\mathcal{F}_1| \) edges out of \( d(v) - |\mathcal{B}| \) possible neighbors.
Thus we use,
\begin{equation*}
T[\mathcal{F}_1](v) = \binom{d(v) - |\mathcal{B}|}{|\mathcal{F}_1|}.
\end{equation*}
For example, for a vertex $a \in \mathbf{S}_{\tau}$, $T[\mathcal{F}_1](a) = \binom{6 - 2}{3} $.

Next consider the term $T[\mathcal{I}_1]$ with a more general pattern \( P_2 \). In this case,  $\mathcal{I}_1 = \{\{u_4, u_5\},\ \{u_5, u_6\}\}$.
To account for these edges, we must estimate the probability \( p_v \) that $|\mathcal{I}_1|$ necessary edges exist among the \( |\mathcal{F}_1| \) neighbors of \( v \). 
One naive way to estimate \( p_v \) is by using the local clustering coefficient \( C_v \) of \( v \)~\cite{clustering_coeff}. Since \( C_v \) represents the probability that two neighbors of \( v \) are connected by an edge, and there are \( |\mathcal{I}_1| \) such edges required, we can estimate \( p_v \) as \( C_v^{|\mathcal{I}_1|} \).

However, naively estimating \( p_v \) as \( C_v^{|\mathcal{I}_1|} \) tends to overemphasize the effect of \( C_v \), leading to inaccurate approximations. This is because real-world graphs tend to form clusters~\cite{graph_cluster_through_mutual_friends,graph_clusters, graph_clusters2, graph_clusters3, graph_clusters4}, and that $|\mathcal{I}_1|$ edges are not independent from each other.
For example, the probability that an edge \( \{a, b\} \) exists given that \( a, b, c \in \mathcal{N}(v) \) and edges \( \{a, c\},\ \{b, c\} \) exist is generally higher than \( C_v \) due to the clustering effect.

Therefore, instead of multiplying \( C_v \) for every edge in \( \mathcal{I}_1 \), we construct an effective subset \( \mathcal{I}_{\text{effective}} \subseteq \mathcal{I}_1 \) of edges that can be considered independent. 
%
For this, we consider the subgraph induced by $\mathcal{I}_1$: \( P_{\text{ind}} = (V_{\text{ind}}, E_{\text{ind}}) \), where \( V_{\text{ind}} = \{ u \mid u \in e \in \mathcal{I}_1 \} \) and \( E_{\text{ind}} = \mathcal{I}_1 \). We then select a spanning forest of \( P_{\text{ind}} \), whose edge set is \( \mathcal{I}_{\text{effective}} \). 
We then estimate
\begin{equation*}
T[\mathcal{I}_1](v) =  C_v^{|\mathcal{I}_{\text{effective}}|}.
\end{equation*}
In the case of pattern \( P_2 \), the induced subgraph \( P_{\text{ind}} \) forms a simple path over three vertices \( u_4, u_5, u_6 \). Since a path is already a tree, the entire edge set \( \mathcal{I}_1 \) serves as its own spanning tree. 
Therefore, for this particular case, we set $ \mathcal{I}_{\text{effective}} = \mathcal{I}_1 = \{\{u_4, u_5\},\ \{u_5, u_6\} \}$. We then have $T[\mathcal{I}_1](a) = C_a^2=(5/15)^2$.

Lastly, \( T[\mathcal{F}_2](v) \) accounts for the two-hop topology of the pattern. For the pattern \( P_3 \), \( \mathcal{F}_2 = \{ \{u_4, u_7\}, \{u_6, u_8\} \} \).  
For each edge in \( \mathcal{F}_2 \), the number of successful extensions is estimated to increase proportionally to the degree of the corresponding vertices.  
For example, \( T[\mathcal{F}_2](v) = d(v_4) \cdot d(v_6) \) appropriately accounts for potential connections to \( u_7 \) and \( u_8 \).
However, since we do not know the exact 1-hop neighbors $(v_4,v_6)$ that will be sampled in the future, we use the average degree of the neighbors of \( v \) for actual computation. 
Thus,
\begin{equation*}
T[\mathcal{F}_2](v) = \left(\frac{\sum_{x \in \mathcal{N}(v)} d(x)}{d(v)}\right)^{|\mathcal{F}_2|}.
\end{equation*}
For example, $T[\mathcal{F}_2](a) = \left(\frac{4+4+3+3+3+3}{6}\right)^2$, considering neighbors of $a$ starting from $v_1$ in a clockwise manner.

We provide a formal definition and an equation to calculate $\mathbf{f_{\text{approx}}}(\cdot \mid \tau)$.

\begin{definition}[Auxiliary Arrays]\label{def:auxiliary_arrays}
Given a pattern \(P\) with matching order 
$
\pi: \{1,\dots,|V_P|\}\;\to\;V_P,
\quad\text{so that}\quad
u_i = \pi(i)\in V_P
$
is the \(i\)-th pattern vertex, we define the \emph{Auxiliary Arrays} \(A^\pi\) as collections of arrays of sets:
\[
  A^\pi = \bigl\{\,\mathcal{B}^\pi,\;\mathcal{F}_1^\pi,\;\mathcal{F}_2^\pi,\;\mathcal{I}_{\mathrm{effective}}^\pi\bigr\},
\]
where each \(\mathcal{B}^\pi,\mathcal{F}_1^\pi,\mathcal{F}_2^\pi,\mathcal{I}_{\mathrm{effective}}^\pi\)
is an array indexed by \(i=1,\dots,|V_P|\), and each element
\(\mathcal{B}^\pi[i],\dots,\mathcal{I}_{\mathrm{effective}}^\pi[i]\)
is the set associated with the vertex \(u_i=\pi(i)\).
\end{definition}

Based on $A^\pi$, the equation for approximating \( n_{\tau \concat v} \), where \( v \) is the candidate for the \( i \)-th vertex \( v_i \), is given by
\begin{equation}
\resizebox{.9\linewidth}{!}{$
n_{\tau \concat v} \approx 
\underbrace{\binom{d(v) - |\mathcal{B}^\pi[i]|}{|\mathcal{F}_1^\pi[i]|}}_{T[\mathcal{F}_1]} 
\times 
\underbrace{C_v^{|\mathcal{I}_{\text{effective}}^\pi[i]|}}_{T[\mathcal{I}_1]}  
\times 
\underbrace{\left( \frac{\sum_{x \in \mathcal{N}(v)} d(x)}{d(v)} \right)^{|\mathcal{F}_2^\pi[i]|}}_{T[\mathcal{F}_2]}.
$}
\label{eq:probability_function}
\end{equation}
We compute this value for all candidate vertices \( v \in \mathbf{S}_\tau \), and use normalized values as the distribution \( \mathbf{f}_{\text{approx}}(\cdot \mid \tau) \).

\subsection{Analysis on Sample Complexity}
\label{sec:method:sample_complexity}

To develop theoretical insight on our sampler proposed in~\cref{sec:approximation}, we analyze the sample complexity.
For each successful trajectory $\tau_i$, the sampler’s output can be written as 
\[
X_{\tau_i} = C(G,P) \times (1 + \eta_i),
\]
where $\eta_i$ denotes the multiplicative error associated with trajectory $\tau_i$ arising from the use of \fapx in place of the ideal distribution \fideal.
Since an unsuccessful trajectory yields \(X_\tau=0\), the second moment of the sampler output is
\begin{align*}
\mathbb{E}[X^{2}]
  &=\sum_{\tau}
      \mathbbm{1}_{\{\tau\text{ successful}\}}\,
      X_\tau^{2}p_\tau
    =\sum_{i=1}^{C(G,P)} X_{\tau_i} \\[2pt]
  &=C(G,P)^{2}\,(1 + \overline{\eta}),
\end{align*}
where $\overline{\eta}$ is the average multiplicative error
\[
\overline{\eta}
  =\frac{1}{C(G,P)}
   \sum_{i=1}^{C(G,P)}\eta_i.
\]
Combining~\cref{thm:unbiased_sampling} with Chebyshev’s inequality, a \((1\!\pm\!\epsilon)\)-relative error is achieved with probability at least \(1-\delta\) whenever
\begin{equation}
N
  \ge
  \frac{\overline{\eta}}{\epsilon^{2}\delta}.
\label{eq:sample-complexity}
\end{equation}
This yields an explicit relationship between $N$ (the number of samplers required), the accuracy of \fapx, and the parameters $(\epsilon,\delta)$. The closer \fapx is to \fideal, the closer \(\overline{\eta}\) is to zero, and the fewer samplers are required.

To illustrate, consider the simplest non-degenerate pattern, the triangle. 
For a successful trajectory \(\tau_i=(v_1,v_2,v_3)\), the multiplicative error $\eta_i$ for the proposed sampler output is
{\small
\begin{equation}
\eta_i
  =\bigg[ \frac{1}{d(v_1)\bigl(d(v_1)-1\bigr)\,C_{v_1}}\displaystyle\sum_{v\in\mathcal{N}(v_1)}\!\bigl(d(v)-1\bigr)
   \cdot
   \frac{\bigl|\mathcal{N}(v_1)\cap\mathcal{N}(v_2)\bigr|} 
         {d(v_2)-1} \bigg] - 1.
\label{eq:alpha-def}
\end{equation}
}
The ratio
\(\frac{|\mathcal{N}(v_1)\cap\mathcal{N}(v_2)|}{d(v_2)-1}\)
approximates the fraction of edges emanating from a neighbor of \(v_1\) that remain inside \(\mathcal{N}(v_1)\).
Because \(\sum_{v\in\mathcal{N}(v_1)}(d(v)-1)\) counts all edges incident to \(\mathcal{N}(v_1)\), the product
\[
\sum_{v\in\mathcal{N}(v_1)}\!\bigl(d(v)-1\bigr)\,
\frac{|\mathcal{N}(v_1)\cap\mathcal{N}(v_2)|}{d(v_2)-1}
\]
serves as a proxy for the number of edges internal to \(\mathcal{N}(v_1)\), whose exact value is \(d(v_1)\bigl(d(v_1)-1\bigr)C_{v_1}\).
Consequently, regardless of the data graph $G$, \cref{eq:alpha-def}, and thus \(\overline{\eta}\), is typically close to zero.

On the other hand, for a sampler based on \texttt{NS-prune} without symmetry breaking, the error is given as
\begin{equation}
\eta_i'
  =\bigg[ \frac{2|E_G|\cdot |\mathcal{N}(v_1)\cap\mathcal{N}(v_2)|}{C(G,P)} \bigg] - 1.
  \label{eq:eta_prime}
\end{equation}
If $G$ is uniform, every edge produces the same number of triangles and \cref{eq:eta_prime} is exactly $0$.
However, as the graph deviates from uniformity, $\eta_i'$ deviates further from 0, and more samplers are required.
Thus, \cref{eq:sample-complexity,eq:alpha-def,eq:eta_prime} suggest that the approximation strategy in~\cref{sec:approximation} can be substantially more effective than existing approaches.

\subsection{Preprocessing Data Graphs}
\label{sec:method:preprocessing}
For a given graph $G$, we 
preprocess three pieces of information, which are used for the calculation of \( \mathbf{f}_{\text{approx}} \) from \cref{eq:probability_function}.

\begin{enumerate}[]
\item The binomial coefficient table \( \binom{n}{k} \) for $n \geq k$.
\item The average neighbor degree information \( \frac{\sum_{x \in \mathcal{N}(v)} d(x)}{d(v)} \).
\item The clustering coefficient \( C_v \).

\end{enumerate}
While the first two items are relatively cheap to process during the graph loading phase, 
computing the exact value of \( C_v \) is time-consuming since it is equivalent to counting all the triangles in the graph. 
Instead, we estimate \( C_v \) by sampling. For each vertex \( v \), we perform sampling \( d(v) \) times and estimate \( C_v \) from the collected data. 
We will discuss the preprocessing time in \cref{sec:eval}. We also note that we bound the estimated clustering coefficient values to a small nonzero constant so that \fapx does not contain any unintended zeros which guarantees the assumption of \cref{thm:unbiased_sampling}.

\section{\thiswork System Design}
\label{sec:pgns}
Based on \fapx, we build the \thiswork system.
We describe how to strategically use \fapx to accelerate the system (\cref{sec:pgns:hybrid}), how to build the matching order (\cref{sec:pgns:order}), followed by the complete procedure (\cref{sec:pgns:procedure}), and system overview with implementation details (\cref{sec:pgns:details}).

\subsection{\SAMPLING}
\label{sec:pgns:hybrid}
The proposed  \fapx can dramatically reduce the required number of samplers for convergence. 
However, naively using $\mathbf{f}_{\text{approx}}$ can sometimes lead to longer execution time compared to using $\mathbf{f}_{\text{uniform}}$. 
While sampling from $\mathbf{f}_{\text{uniform}}$ can be done in $O(1)$ time, sampling from $\mathbf{f}_{\text{approx}}$ takes $O(|V_G|)$ time for constructing the distribution. 
Even though the number of samplers is still significantly smaller than that of using \funi, the overall latency can be longer.

To address this problem, the proposed \emph{\sampling} comprises a decision heuristic that determines the type of sampling distribution for each vertex based on the structure of the remaining unsampled subgraph of $P$. 
In the decision heuristic, we consider the density of the remaining unsampled subpattern.
We find that as the unsampled portion of the pattern gets denser, $n_{\tau \concat v}$ becomes less predictable, and utilizing $\mathbf{f}_{\text{uniform}}$ is more beneficial.

Using the difference between the number of edges and the number of vertices to represent density, we define the following decision function: 

\begin{definition}[Decision Function]\label{def:decision_function}
Given a pattern \(P\), a matching order 
$
  \pi: \{1,\dots,|V_P|\}\;\to\;V_P,
  \quad\text{so that}\quad
  u_i=\pi(i) \in V_P,
$
and a threshold \(\beta\le1\), the decision function
\(\mathcal{D}^\pi:\{1,\dots,|V_P|\}\to\{0,1\}\) is defined as follows:

\begin{enumerate}[leftmargin=5.0mm]
    \item \textbf{Modeling the Certainty of Using $\mathbf{f}_{\text{approx}}$}:
    \begin{itemize}[leftmargin=.5mm]
        \item Define the set of vertices at least two hops away from \( u_i \) in \( P_i=(V_i, E_i) \subseteq P \) (\cref{eq:Pi}):
        \[
        V^{\geq 2\text{hop}}_i = \{ u \in V_i \mid 2 \leq \text{dist}_{P_{i}}(u_i, u) < \infty \}.
        \]
        \item Define the set of edges connected to \( V^{\geq 2\text{hop}}_i \):
        \[
        E^{\geq 2\text{hop}}_i = \left\{ \{u_a, u_b\} \in E_P \,\middle|\, u_a \in V^{\geq 2\text{hop}}_i \text{ or \ } u_b \in V^{\geq 2\text{hop}}_i \right\}.
        \]
        \item Define the certainty to use \( \mathbf{f}_{\text{approx}} \):
        \[
        \text{Certainty}(i) = 1 - \frac{|E^{\geq 2\text{hop}}_i| - |V^{\geq 2\text{hop}}_i|}{|V_i|}.
        \]
     The value is bounded above by one, i.e., $\text{Certainty}(i) \leq 1$.
    \end{itemize}
    \item \textbf{Decision Function}:
    \[
    \mathcal{D}^\pi(i) =  \mathbbm{1}_{\left( \text{Certainty}(i) \geq \beta \right)}
    \]
\end{enumerate}
\label{def:decision_function}
\end{definition}
Here, we use $\beta=0.8$ as the default value. \( \mathcal{D}^\pi(i) = 1 \) indicates that we use \( \mathbf{f}_{\text{approx}} \) for sampling \( v_i \), and \( \mathcal{D}^\pi(i) = 0 \) means we use \( \mathbf{f}_{\text{uniform}} \).
This allows us to 
combine the strengths of both distributions based on the topology of the remaining subpattern.
Note that this can be precomputed only from $P$ with negligible cost. 

\subsection{Matching Order Construction} 
\label{sec:pgns:order}
Since the decision function \( \mathcal{D}^\pi \) depends on the matching order \( \pi \), it is crucial to construct \( \pi \) wisely to maximize the use of \( \mathbf{f}_{\text{approx}} \). To achieve this, we propose an algorithm that incrementally builds the matching order in a greedy manner, as detailed in Algorithm~\ref{alg:matching_order}.

\begin{algorithm}
\caption{Matching Order Construction}\label{alg:matching_order}
\begin{algorithmic}[1]
\State Initialize empty list \( \pi \gets [\,] \)

\For{$i$ in $[1, |V_P|]$}
    \If{\( i=1 \)}
        \State \( \mathcal{N} \gets V_P \)
    \Else
        \State \( \mathcal{N} \gets \{ u \in V_P \setminus \pi \mid \exists u' \in \pi \text{ such that } \{u, u'\} \in E_P \} \)
    \EndIf
    \For{\( u \in \mathcal{N} \)}
        \State $\pi_u \gets \pi+u$, \( s(u) \gets |\mathcal{F}_1^{\pi_u}[i]| + |\mathcal{I}_1^{\pi_u}[i]|\)
    \EndFor
    \State \( \pi \gets \pi + \operatorname*{argmax}_{u \in \mathcal{N}} s(u) \)

\EndFor
\State \Return \( \pi \)
\end{algorithmic}
\end{algorithm}

As shown in Lines~3--7, we first construct a candidate set $\mathcal{N}$ of neighboring vertices so that we always explore the pattern in a connected fashion.
%
%
In Lines~8--10, we calculate the score $s(u)$ for all \( u \in \mathcal{N} \), where \( |\mathcal{F}_1^{\pi_u}[i]|\) and \( |\mathcal{I}_1^{\pi_u}[i]| \) are defined as in \cref{sec:approximation}, assuming that vertex \( u \in V_P \) is selected as the next vertex in the matching order. In Line~11, by selecting the vertex that maximizes this sum, we prioritize vertices with the largest unsampled 1-hop structures. This approach results in \( \mathbf{f}_{\text{approx}} \) being selected more frequently, as the \textit{Certainty} value in \cref{def:decision_function} will be higher.



\begin{algorithm}[t]
\caption{Structure-Informed Neighbor Sampling}\label{alg:PGNS}
\begin{algorithmic}[1]

\Procedure{Sample}{$G$, $P$, $\pi$, $\mathcal{D}^\pi$, $\mathcal{A}^\pi$} 
    \State Initialize $\tau \gets []$, $p \gets 1.0$
    \For{$i$ in $[1, |V_P|]$}
        \State \textit{$\triangleright$ \ Build set $\mathbf{S}_\tau$ of candidate vertices}
        \If{$i = 1$}
            \State $\mathbf{S}_\tau \gets V_G$
        \Else
            \State $\mathbf{S}_{\tau} \gets \{v\mid v \notin \tau, v\in \bigcap_{ \{u_j,u_i\} \in E_P , j<i} \mathcal{N}(v_j)\}$ 
        \EndIf
        
        \State \textbf{if $|\mathbf{S}_\tau| = 0$ then} \Return 0
        
        \State \textit{$\triangleright$ \ Build probability distribution}
        \If{ $\mathcal{D}^\pi(i) = 1$}
        \State calculate $\mathbf{f_{\text{approx}}}$ using $(\mathbf{S}_\tau,\mathcal{A}^\pi, i)$ with \cref{eq:probability_function}
        \State $\mathbf{f} \gets \mathbf{f_{\text{approx}}}$
        \Else
        \State $\mathbf{f} \gets \mathbf{f}_{\text{uniform}}$
        \EndIf
        \State $v \gets rand\_select(\mathbf{S}_\tau$, $\mathbf{f})$
        \State $\tau \gets \tau + v, p = p \cdot \mathbf{f}(v)$
    \EndFor
    \State \Return  $\frac{1}{p}$ 
\EndProcedure

\end{algorithmic}
\end{algorithm}

\subsection{Complete Sampling Procedure}
\label{sec:pgns:procedure}
The procedure of the sampling algorithm used in \thiswork is displayed in \cref{alg:PGNS}.
In Lines~4--9, we construct the candidate set \( S_\tau \) using connectivity constraints. If no possible vertices exist (Lines~10), we return a value of 0 and terminate the sampling process. Otherwise, we build a probability distribution function according to the decision function \( \mathcal{D}^\pi \) (Lines~11--17). We then sample one vertex from \( S_\tau \) following this distribution and update the probability \( p \) accordingly (Lines~18--19). By repeating this vertex sampling process \( |V_P| \) times, we return \( 1/p \) as the sampler output.

\begin{figure}[t]
    \centering
    \includegraphics[width=0.95\columnwidth]{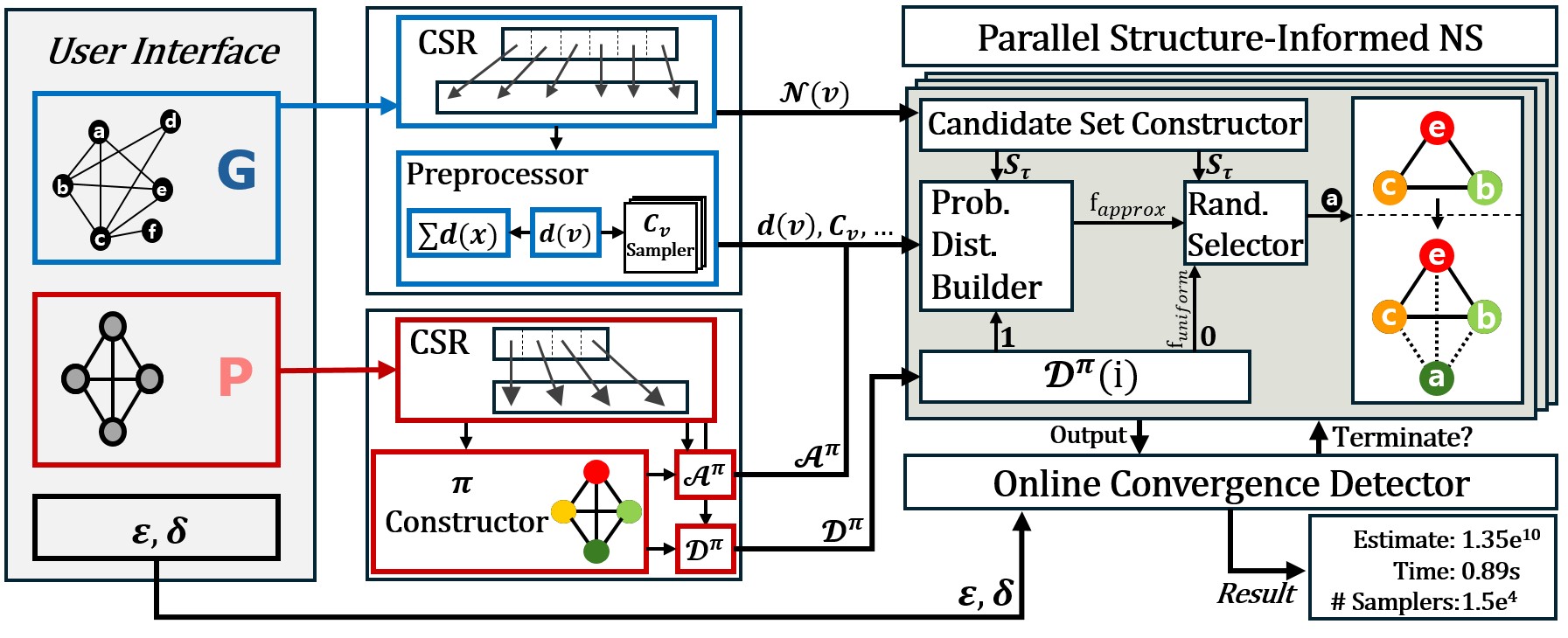} 
    \caption{Overview of \thiswork System.}
    \label{fig:systemoverview}
\end{figure}

We note that sampling solely based on \( \mathbf{f}_{\text{uniform}} \) is equivalent to ScaleGPM's \texttt{NS-prune} procedure without applying symmetry breaking.
The definition of \( \mathbf{f}_{\text{uniform}} \) varies depending on step $i$. 
For \( i = 1 \) (sampling the first vertex), \( \mathbf{f}_{\text{uniform}} \) is \emph{edge-uniform}, meaning that vertices are sampled with probability proportional to their degrees. For \( i \geq 2 \), \( \mathbf{f}_{\text{uniform}} \) is \emph{vertex-uniform}, so each vertex in \( \mathbf{S}_\tau \) is sampled with equal probability.



\begin{figure*}[t]
    \centering
    \begin{minipage}[t]{0.48\textwidth}
        \centering
        \vspace{-28mm}
        
    \footnotesize  

    \renewcommand{\arraystretch}{1.2}
    \begin{tabular}{lrrr}
    \\ 
        \toprule
        \textbf{Graph} &\textbf{$|V_G|$} & \textbf{$|E_G|$} & Max Degree \\
        \midrule
        LiveJournal (\texttt{Lj}) \cite{lj, snap} & $4.8\times 10^6$ & $4.3\times 10^7$ & $2.0\times 10^4$ \\
        Uk-2002 (\texttt{Uk}) \cite{uk,law1, law2}& $1.8\times10^7$ & $2.6\times10^8$ & $1.9\times10^5$ \\
        Twitter (\texttt{Tw}) \cite{twitter, snap} & $4.2\times10^7$ & $1.2\times10^9$& $3.0\times10^6$  \\
        Friendster (\texttt{Fs}) \cite{fs, snap} & $6.6\times10^7$ & $1.8\times10^9$ & $5.2\times10^3$ \\
        Gsh-2015 (\texttt{Gsh}) \cite{law1, law2} & $9.9\times10^8$ & $2.6\times10^{10}$ & $5.9\times10^7$ \\
        \bottomrule \vspace{-.99mm}
    \end{tabular}
        \captionof{table}{Graph datasets used for evaluation. }
        \label{tab:graph-datasets}
    \end{minipage}
    \begin{minipage}[t]{0.48\textwidth}
        \centering
        \includegraphics[width=.7\columnwidth]{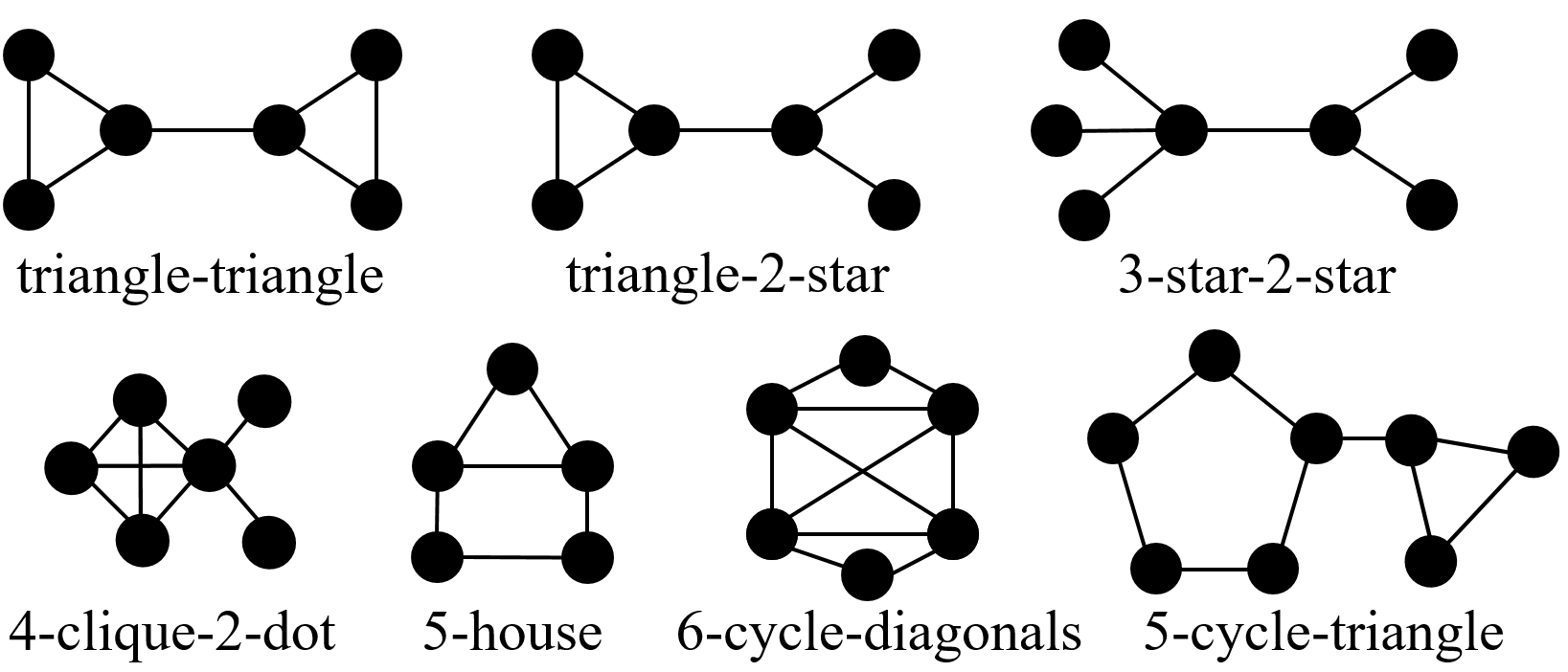} 
        \caption{Patterns used for evaluation.}
        \label{fig:pattern}
    \end{minipage}%
\end{figure*}

\begin{table*}[ht]
    \centering
    \footnotesize

    \renewcommand{\arraystretch}{1.2}
    \begin{tabular}{l cccc cccc cccc }

        \hline 
        
         \multirowcell{2}[-4pt][l]{\textsc{Instance}} & 
        \multicolumn{4}{c}{4-clique} & \multicolumn{4}{c}{4-clique-2-dot} & \multicolumn{4}{c}{6-clique}  \\
        
    \cmidrule(lr){2-5}\cmidrule(lr){6-9}\cmidrule(lr){10-13}

        & \texttt{Lj} & \texttt{Uk} & \texttt{Tw} & \texttt{Fs}
        & \texttt{Lj} & \texttt{Uk} & \texttt{Tw} & \texttt{Fs}
        & \texttt{Lj} & \texttt{Uk} & \texttt{Tw} & \texttt{Fs}
      \\

        \hline
        \multirow{1}{*}{\textsc{\thiswork}} 
        &\textbf{0.002}&	\textbf{0.006}&	\textbf{0.016}&	0.028&	
        \textbf{0.003}&	\textbf{0.069}&	\textbf{5.330}&	\textbf{0.045}& 
        \textbf{0.001}&	\textbf{0.005}&	\textbf{0.017}&	\textbf{0.143}\\

        \multirow{1}{*}{\textsc{ScaleGPM}} 
        &0.007 &	0.014&	0.018	&\textbf{0.005}&	0.052	&191.9&	1930 &	0.089&
        0.095	&0.843&	0.377	&2.460\\
        
        \multirow{1}{*}{Arya} 
         & 504.4 &  688.6 & TO & TO & 
         2126 & 1528 & TO & TO&
         TO & TO & TO & TO \\

        \multirow{1}{*}{ASAP} 
         & 958.2 &  462.4 & TO & TO & 
         TO & TO & TO & TO & 
         TO & TO & TO & TO \\
        
        \hline
        \multirow{1}{*}{Peregrine} 
        & 6.779 &  128.0 & TO & 199.5 & TO & TO & TO & TO & TO & TO & TO & 1114  \\

        \Xhline{1.5pt}

        \multirowcell{2}[-4pt][l]{\textsc{Instance}}  & \multicolumn{4}{c}{3-star-2-star} & \multicolumn{4}{c}{triangle-2-star} & \multicolumn{4}{c}{triangle-triangle}\\
        
    \cmidrule(lr){2-5}\cmidrule(lr){6-9}\cmidrule(lr){10-13}
    
        & \texttt{Lj} & \texttt{Uk} & \texttt{Tw} & \texttt{Fs}
        & \texttt{Lj} & \texttt{Uk} & \texttt{Tw} & \texttt{Fs}
        & \texttt{Lj} & \texttt{Uk} & \texttt{Tw} & \texttt{Fs}
         \\

        \hline
        \multirow{1}{*}{\textsc{\thiswork}} 
        &\textbf{0.002}&	\textbf{0.024}&	\textbf{0.840}&	\textbf{0.028}&	
        \textbf{0.002}&	\textbf{0.081}&	\textbf{4.701}&	\textbf{0.030}&	
        \textbf{0.001}&	\textbf{0.013}&	\textbf{4.738}&	\textbf{0.044} \\
        \multirow{1}{*}{\textsc{ScaleGPM}}         
        & 404.1&TO&	TO& 	0.054&0.197& 40.53& 5009& 0.043& 	0.025& 0.186& 33.75& 0.096 \\

        \multirow{1}{*}{Arya} 
         & 216.4 &  1345 & 154.1 & 135.4 & 
         140.2 & 177.2 & 632.1 & 1640 & 
         686.3 & 1827 & 9148 & TO  \\

        \multirow{1}{*}{ASAP} 
         & 2372 & TO & TO & 3.067
         & TO & TO & TO & 6273
         & 4628 & TO & TO & TO  \\

        \hline 
        \multirow{1}{*}{Peregrine} 
         & TO &  TO & TO & TO & TO & TO & TO & TO & TO & TO & TO & TO\\

        \hline
    \end{tabular}
    \vspace{1mm}
    \caption{Execution time (sec.) for patterns where $\mathcal{D}^\pi = 1$. TO: timed out. Fastest time in bold.}
    \label{tab:main}
\end{table*} 



\subsection{System Overview}
\label{sec:pgns:details}
Based on \sampling, we propose \thiswork, a system for fast approximate graph pattern mining.
The high-level sketch of \thiswork is given in Figure~\ref{fig:systemoverview}.
 It first receives four inputs from the user: (1) a graph file $G$, (2) a pattern file $P$, (3) \(\epsilon\), and (4) \(\delta\). 
 The graph is read and loaded in CSR format. 
 If the graph has not been preprocessed, the preprocessor calculates additional information (\cref{sec:method:preprocessing}).
For the pattern, the system first constructs the matching order \(\pi\) which is then used to calculate \(\mathcal{D}^\pi\) and \(\mathcal{A}^\pi\).

Once the information is ready, the sampling engine is run in parallel, where each thread is assigned an individual sampler. 
For every fixed number of iterations, the sampler outputs are checked for convergence by the online convergence detector (\cref{alg:convergences}) proposed by ScaleGPM
,
utilizing \(\epsilon\) and \(\delta\).
Finally, the results, such as the estimated pattern count, are output to the user.

When the queried pattern $P$ is a clique, a popular approach is to apply orientation optimization~\cite{orientation1, g2miner, orientation2, orientation3}. 
It converts the undirected graph into a directed acyclic graph (DAG) by removing edges pointing from high-degree vertex to low-degree vertex.
Similar to ScaleGPM \cite{scalegpm}, which utilizes this method, we apply the same optimization for clique patterns.

\section{Evaluation}
\label{sec:eval}
\thiswork is implemented in C++ with OpenMP for multithreading. 
We evaluated \thiswork over a diverse set of data graphs and patterns. 
We used real-world graphs spanning a wide range of sizes, as summarized in Table~\ref{tab:graph-datasets}. 
For the patterns to be mined, we selected a diverse set of patterns, partly drawn from prior works~\cite{arya, scalegpm}, as illustrated in \cref{fig:pattern}.
We compared \thiswork against the state-of-the-art AGPM systems ASAP~\cite{asap}, Arya~\cite{arya} and ScaleGPM~\cite{scalegpm} and an exact mining system Peregrine~\cite{peregrine}. 
We used the official implementations for Peregrine, and Arya.
We faithfully reproduced ASAP from scratch as their codes are not publicly available.
Because the official ScaleGPM implementation requires custom implementation for individual patterns, 
we implemented a general-purpose version of ScaleGPM and, where applicable, reported the minimum execution time observed between our implementation and the official code.

\begin{figure*}[t] 
    \begin{minipage}{0.64\textwidth} 
        \centering

\renewcommand{\arraystretch}{1.2}

    \footnotesize

    \begin{tabular}{l cc  cc  cc }

        \hline 
        
        \multirowcell{2}[-4pt][l]{\textsc{Instance}}  & \multicolumn{2}{c}{5-house} & \multicolumn{2}{c}{6-cycle-diagonals} & \multicolumn{2}{c}{5-cycle-triangle} \\

        \cmidrule(lr){2-3} \cmidrule(lr){4-5} \cmidrule(lr){6-7}
        & \texttt{Lj}  & \texttt{Tw} 
        & \texttt{Lj}  & \texttt{Tw}
        & \texttt{Lj}  & \texttt{Tw}  \\

        \hline
        \multirow{1}{*}{\textsc{\thiswork}} 
        &\textbf{0.003}&	\textbf{2.775}&	\textbf{0.003}&	\textbf{0.580}&	\textbf{0.036}& \textbf{188.9} \\	
        \multirow{1}{*}{\textsc{ScaleGPM}}         
        & 0.009	&31.72	&0.022&	27.25&	0.041&	2002 \\
        
        \multirow{1}{*}{Arya} 
         & 65.83 &  5490 & TO & TO & TO & TO  \\

        \multirow{1}{*}{ASAP} 
         & 117.8 &  TO 
         & TO & TO 
         & TO & TO \\
        
        \hline
        \multirow{1}{*}{Peregrine} 
        & 669.1 &  TO & 1483 & TO & TO & TO \\

        \hline
    \end{tabular}
        \captionof{table}{Execution time (sec.) for patterns where $\mathcal{D}^\pi \neq 1$. TO: timed out. } 
        \label{tab:typeB}
    \end{minipage}%
    \setlength{\abovecaptionskip}{1pt} 
    \setlength{\belowcaptionskip}{0pt} 
    \begin{minipage}{0.34\textwidth} 
        \centering
        \includegraphics[height=3cm, width=\linewidth]{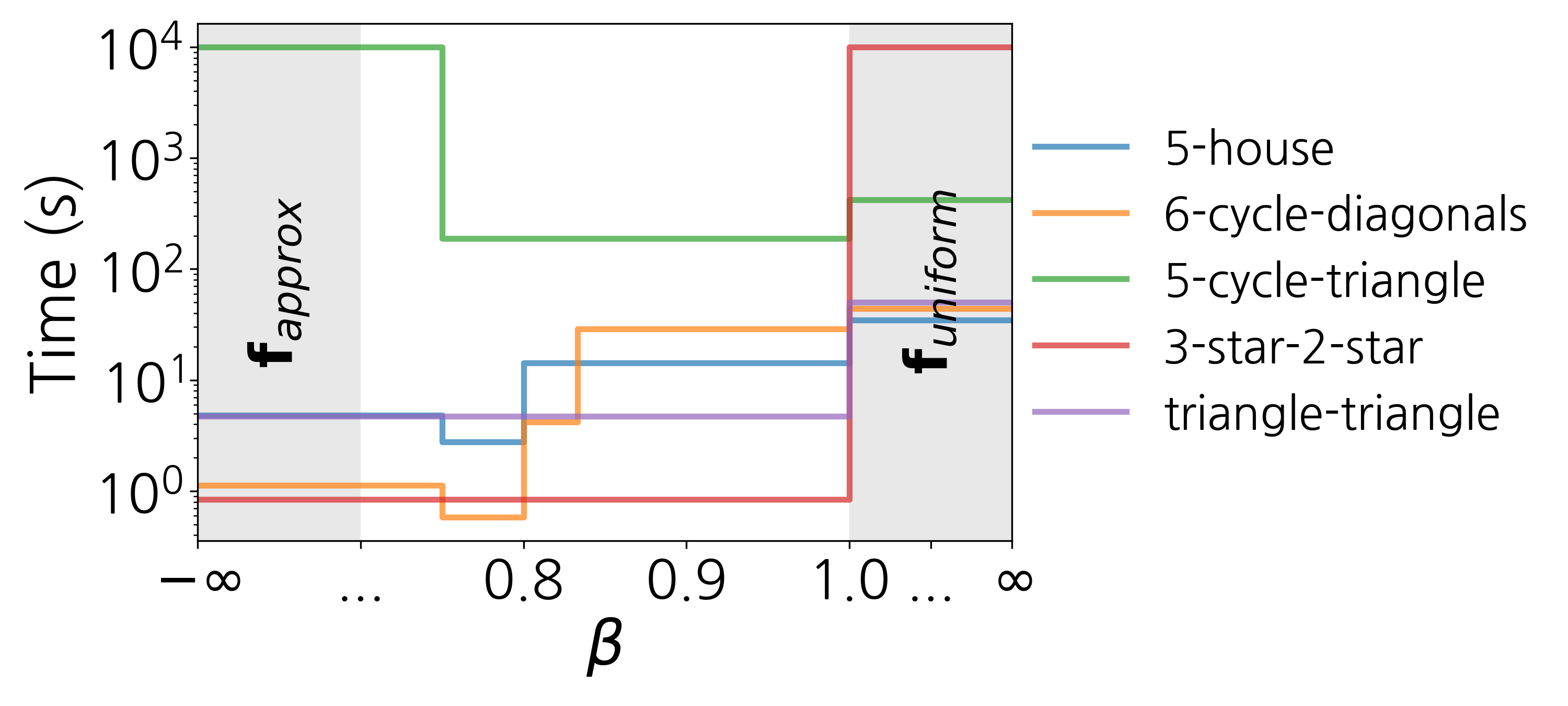} 
        \caption{Impact of $\beta$ on execution time over \texttt{Tw} graph.}
        \label{fig:beta}
    \end{minipage}
\end{figure*}

\begin{table*}[t]
    \centering
    \footnotesize

    \begin{subtable}{0.49\textwidth}
        \centering
        
        


        
        

    \renewcommand{\arraystretch}{1.2}

    \begin{tabular}{l ccc  }
        
        \hline
        
         \multicolumn{1}{l}{\textsc{Pattern}}  & 4-clique & 5-house & triangle-triangle  \\

        \hline
        \multirow{1}{*}{\textsc{\thiswork}} 
         & 0.443 &  \textbf{89.11} & \textbf{1120}    \\
        
        \multirow{1}{*}{\textsc{ScaleGPM}}     
         & \textbf{0.130} &  731.1 &  TO    \\
        
        \multirow{1}{*}{Arya} 
         & OoM & OoM & OoM    \\

        \multirow{1}{*}{ASAP} 
         & TO &  TO & TO   \\

        \hline
        \multirow{1}{*}{Peregrine} 
         & TO & TO & TO \\
        \hline

    \end{tabular}

        \caption{Execution time on \texttt{Gsh} graph.}
        \label{tab:huge}
    \end{subtable}
    \hspace{0.01\textwidth} 
    \begin{subtable}{0.49\textwidth}
        \centering
        
\renewcommand{\arraystretch}{1.2}

\begin{tabular}{l ccc ccc}
    
    \hline
    \multicolumn{1}{c}{}
    & \multicolumn{3}{c}{\texttt{Lj}, 4-clique} & \multicolumn{3}{c}{\texttt{Lj}, 5-house} \\

     \cmidrule(lr){2-4}
\cmidrule(lr){5-7}
    \multicolumn{1}{l}{Error bound $\epsilon$}
     & 10$\%$ & 1$\%$ & 0.1$\%$ 
    & 10$\%$ & 1$\%$ & 0.1$\%$ \\

    \hline
    \multirow{1}{*}{\textsc{\thiswork}} 
     & \textbf{0.002}  & \textbf{0.008} & \textbf{0.528} & 
     \textbf{0.003}  & \textbf{0.026} &  \textbf{2.314} \\
    \multirow{1}{*}{\textsc{ScaleGPM}}    
     & 0.008 & 0.689 & 68.86&
     0.012  & 1.240 &  129.0\\
    \multirow{1}{*}{Arya} 
     & 504.4 &  8545 & TO 
     & 65.83 &  1037  & 3910 \\

     \multirow{1}{*}{ASAP} 
     & 958.2 & 1901 & TO 
     & 117.8 & 573.6 & TO\\

    \hline

\end{tabular}
        \caption{Execution time for different error bounds.}
        \label{tab:error}
    \end{subtable}
    \caption{Execution time (sec.) for (a) huge graph and (b) different error bounds. TO: timed out. Fastest time in bold.}
    \label{tab:two_tables}
\end{table*}

\begin{figure}[t]
    \centering
    \includegraphics[width=0.99\columnwidth]{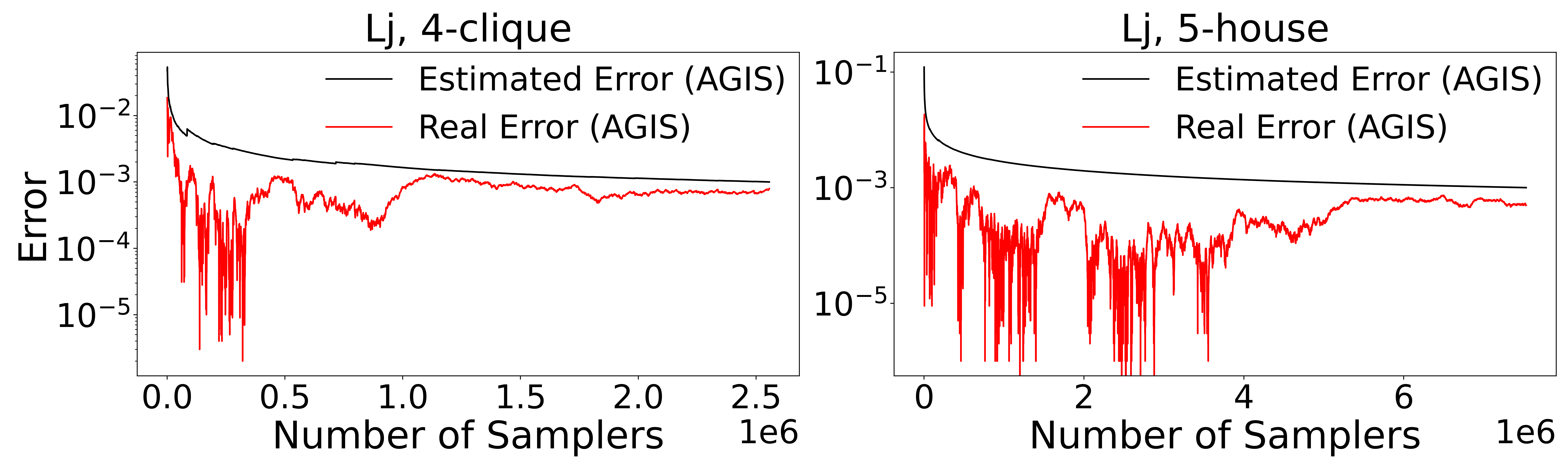} 
    \caption{Real error compared to the estimated error from online convergence algorithm on \thiswork.}
    \label{fig:real vs estimated error}
\end{figure}

For all systems, we excluded the time required to load the input graphs. We also excluded the time spent on pattern-specific preprocessing steps such as solving linear programs for pattern decomposition in Arya, constructing automorphisms and restriction sets for ScaleGPM, and computing $\mathcal{A}^\pi, \mathcal{D}^\pi$ in \thiswork. 
For ScaleGPM, we used only the strict mode (default) of ScaleGPM. This is because the loose mode of ScaleGPM uses graph sparsification with exact counting, which does not use online convergence and thus cannot provide results with high confidence.

To validate the estimated values, we compared the results against Peregrine for all (graph, pattern) pairs that Peregrine can mine within a moderate time. For pairs where Peregrine fails to produce an exact count, the ground truth is unknown. However, for all of the cases, we confirmed that ScaleGPM and \thiswork produce similar estimates, ensuring that $\mid(X_s - X_a)/((X_s+X_a)/2)\mid < 2\epsilon$, where $X_s$ and $X_a$ are the results from ScaleGPM and \thiswork, respectively.

For all baselines and \thiswork, the experiments were conducted on an AMD Ryzen Threadripper PRO 7985WX with 48 physical cores with 512GB DRAM. We used a timeout of \qty{1.0e4}{\second} for all runs.

\subsection{Overall Performance}
\label{sec:eval:overall}

\cref{tab:main} and \cref{tab:typeB} show the main results for diverse patterns. We follow the most common practice of $\epsilon=0.1, \delta=0.01$ if not specified otherwise.
Overall, \thiswork achieves the fastest speed for the majority of patterns and successfully mines all patterns without encountering timeouts.
Specifically, \thiswork obtains 
28.5
$\times$ speedup (using $10^4$s for timeouts) against ScaleGPM, 
and 63,108$\times$ speedup against Peregrine in geometric mean. 


As expected, Peregrine, the exact GPM system fails to produce results within the given time limit for most cases. For graphs larger than \texttt{Lj}, Peregrine can only mine clique patterns where heavy pruning can be applied. This clearly demonstrates the need for AGPM systems over exact mining systems.


Among the baselines, ScaleGPM achieves the best overall performance.
We also observe that ScaleGPM performs the best for clique patterns due to the benefits of DAG orientation optimization.
However, ScaleGPM requires a long time to converge for some (graph pattern) pairs. 
For instance, a 3-star-2-star pattern on \texttt{Lj}  takes more than 100,000$\times$ longer to converge compared to \thiswork.
This observation aligns with those in \cref{sec:limitations}, where patterns containing bridges pose significant challenges under skewed degree distributions.

We can further observe this phenomenon by comparing results on the \texttt{Fs} graph. \texttt{Fs} is distinctive due to its near-uniform degree distribution. As shown in \cref{tab:graph-datasets}, \texttt{Fs} contains more edges than \texttt{Tw}, yet its maximum degree is only about 0.2\% of that of \texttt{Tw}. Since the degree distribution in \texttt{Fs} is not highly skewed, ScaleGPM effectively handles 3-star-2-star and triangle-2-star patterns. Additionally, from a more fundamental standpoint, \funi takes an unintended advantage on \texttt{Fs}, since \funi will be closer to \fideal.

\sloppy{
Furthermore, we can understand why Arya outperforms ScaleGPM on such patterns for highly skewed graphs through a similar reasoning process.}
This is because, unlike ScaleGPM, Arya attempts to achieve convergence on a subsampled graph, thereby effectively reducing the range of its degree distribution.
However, as noted in \cref{sec:back:convergence}, ELP provides no convergence guarantees and the results of the ELP heuristic are observed to be highly unstable. 

\begin{table*}[t]
    \centering
    \footnotesize
    \renewcommand{\arraystretch}{1.2}
    \begin{tabular}{l cc cc  cc cc cc cc }
        \hline 
        \multirowcell{2}[-1.5pt][l]{\textsc{Instance}}  & \multicolumn{2}{c}{4-clique} & \multicolumn{2}{c}{4-cliqe-2-dot} & \multicolumn{2}{c}{6-clique} & \multicolumn{2}{c}{3-star-2-star} & \multicolumn{2}{c}{triangle-2-star} & \multicolumn{2}{c}{triangle-triangle} \\

        \cmidrule(lr){2-3} \cmidrule(lr){4-5} \cmidrule(lr){6-7} \cmidrule(lr){8-9}
        \cmidrule(lr){10-11} \cmidrule(lr){12-13}

         &\texttt{Lj} & \texttt{Tw} & \texttt{Lj} & \texttt{Tw}
        & \texttt{Lj} & \texttt{Tw} & \texttt{Lj} & \texttt{Tw}
        & \texttt{Lj} & \texttt{Tw} & \texttt{Lj} & \texttt{Tw}
       \\

        \hline
        \multirow{1}{*}{\textsc{\thiswork (Structure-Informed NS)}} 
	&\textbf{5.0e2}&		\textbf{2.3e3}&		
    \textbf{5.0e2}&		\textbf{4.0e3}&
    \textbf{7.5e2}&		\textbf{1.0e3}&			
	\textbf{5.0e2}&		\textbf{5.0e2}&		
    \textbf{2.8e3}&		\textbf{1.4e4}&		
    \textbf{5.0e2}&		\textbf{1.1e4} \\        
        \multirow{1}{*}{\textsc{ScaleGPM (NS-prune)}}         
	&5.7e4 &1.3e5&	
    2.9e5& 8.2e7&
    1.2e6& 2.8e6&		
    1.5e9&	\texttt{X}	&	
    8.8e5	&	5.6e7	&	
    1.5e5	&	2.3e6 \\	

        \hline

    \end{tabular}
    \caption{Number of samplers needed for convergence. \texttt{X} for failure in convergence. The smallest number is in bold.}
    \label{tab:numberofsamplersneeded}
\end{table*}
\begin{figure*}[t]
  \centering
  \begin{subfigure}[b]{0.48\textwidth}
    \centering
    \includegraphics[width=\linewidth]{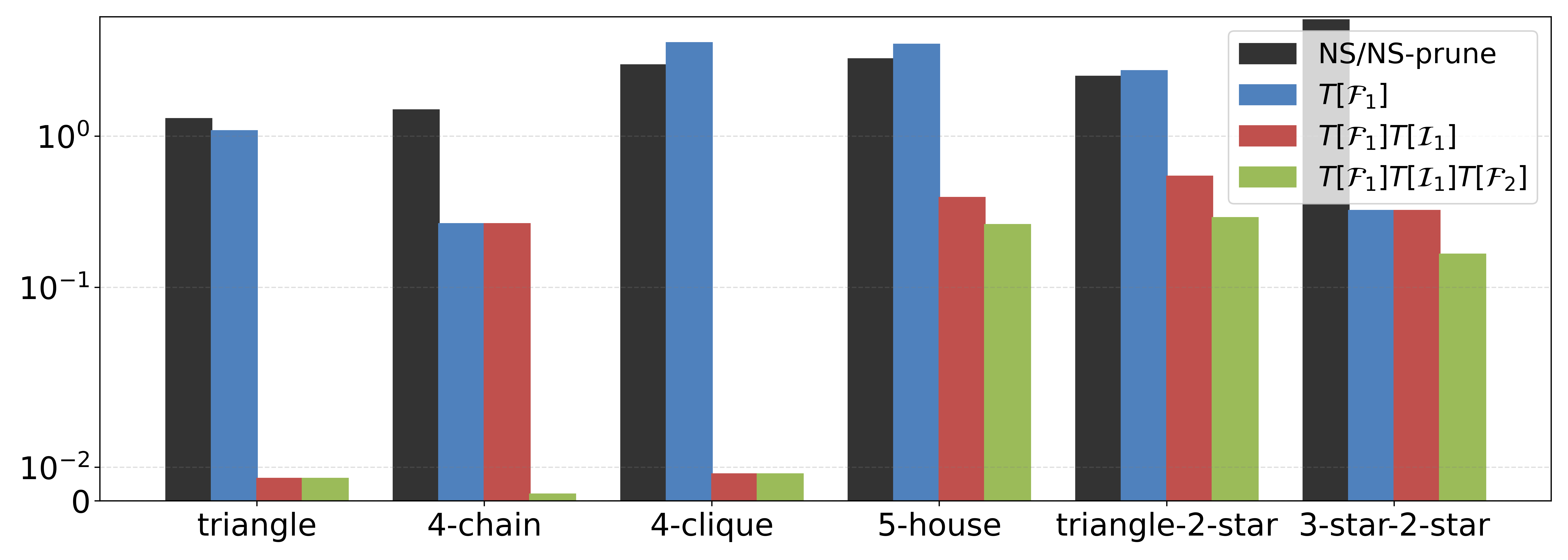}
    \caption{
    KL divergence from each distribution (\fapx and \funi) to the ideal distribution \fideal.}
    \label{fig:fapprox_performance_left}
  \end{subfigure}
  \hfill
  \begin{subfigure}[b]{0.48\textwidth}
    \centering
     \includegraphics[width=\linewidth]{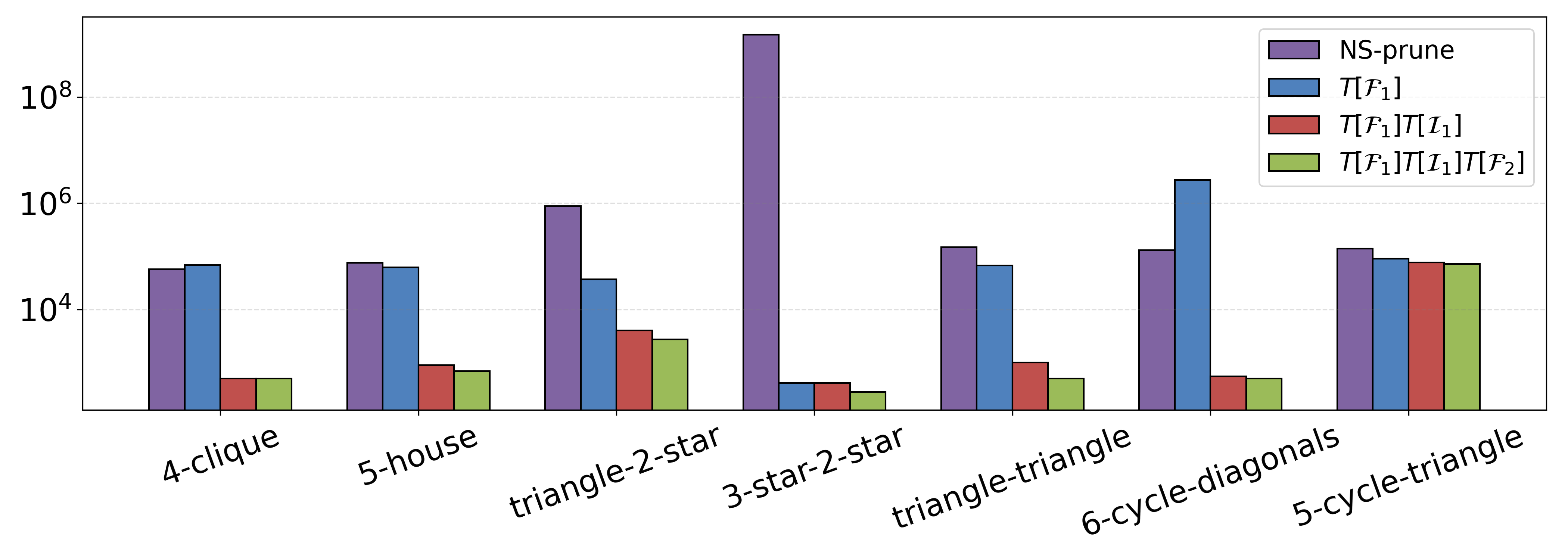}
    \caption{    
Number of samplers needed for convergence.}
    \label{fig:fapprox_performance_right}
  \end{subfigure}
  \caption{Efficiency of the proposed approximation technique. Tested on \texttt{Lj} graph.}
  \label{fig:fapprox_performance}
\end{figure*}

The patterns are grouped into two tables. 
For the patterns examined in \cref{tab:main}, the certainty values (\cref{def:decision_function}) are always maximal (equal to 1). Thus, regardless of which $\beta \leq 1$ value is chosen, \thiswork consistently uses \fapx. 
In contrast, for the patterns in \cref{tab:typeB}, the certainty values are not consistently equal to 1. 
Specifically, in the 5‑house pattern one of five vertices is sampled uniformly; in the 6‑cycle‑diagonals one of six; and in the 5‑cycle‑triangle three of eight.
By testing for numerous patterns including those in \cref{fig:beta}, we find that selecting a threshold value of $\frac{4}{5} \leq \beta \leq \frac{5}{6}$ yields robust and strong performance across a variety of patterns. 
\cref{fig:beta} illustrates how $\beta$ affects the performance. 
Here, $\beta = -\infty$ indicates that \fapx is always used, whereas $\beta=1$ denotes the most conservative use of \fapx. 
For experimental purposes, we extend the range of $\beta$ to include values where $\beta > 1$, which is equivalent to always choosing \funi.
As observed, employing \fapx ($\beta = -\infty$) generally results in faster convergence compared to using \funi ($\beta = +\infty$). Nevertheless, every pattern achieves its optimal performance for some value of $\beta$ within the range $-\infty < \beta \leq 1$, indicating that an optimal point exists between the extremes of exclusively utilizing \fapx and exclusively utilizing \funi. 
By utilizing the proposed $\mathcal{D}^\pi$, \thiswork effectively balances the use of \fapx and \funi.

\cref{tab:huge} presents the performance results on the massive graph \texttt{Gsh}, containing more than 25 billion edges. As shown, only \thiswork successfully scales to this large size. 
ScaleGPM is faster solely for the 4-clique pattern, where the overhead of building \fapx outweighs the complexity of the pattern.
In addition, \cref{tab:error} reports the performance under different error bounds. As indicated, \thiswork is the only framework that can handle $\epsilon = 10^{-3}$ within a few seconds. 
We also verified that the returned counts respect the given $\epsilon$. In \cref{fig:real vs estimated error}, we plot the estimated error provided by the online convergence detector against the actual error, measured using exact mining results. For error levels up to $10^{-3}$, the real error remains bounded by the estimated error, confirming that the online convergence method functions effectively within \thiswork.

We note that there are a few corner cases where \thiswork is slower than ScaleGPM (4‑clique on the \texttt{Fs}, \texttt{Gsh}). The upfront cost of computing the full‑graph vertex sampling distribution can outweigh its benefits when the target pattern is small and the ideal distribution is nearly uniform, rendering the sophisticated setup unnecessary.  


\subsection{Analysis on the Approximate Distribution}
\label{sec:eval:fapprox}

We now examine our estimation technique for the sampling distribution in greater detail. \cref{tab:numberofsamplersneeded} compares the number of samplers required for convergence between \thiswork and ScaleGPM, both of which employ the same convergence detection algorithm. As shown, \thiswork requires orders of magnitude fewer samplers. For instance, ScaleGPM demands three million times more samplers to achieve convergence for the 3-star-2-star pattern in \texttt{Lj}. This highlights that the fundamental speedup of \thiswork arises from leveraging a well-calculated \fapx rather than relying on a naive \funi.

\cref{fig:fapprox_performance} demonstrates how \fapx improves as we incorporate additional terms. In \cref{fig:fapprox_performance_left}, we first count the actual value of $n_{(v)}$ for every vertex $v \in V_G$. With this information, we calculate the ideal distribution for sampling the first vertex of the pattern, $\mathbf{f}_{ideal}(v|())$. We then measure the KL divergence of \fideal for several other distributions: \funi of \texttt{NS}/\texttt{NS-prune}, and \fapx of \thiswork. We analyze three incremental versions of \fapx: (1) using only $T[\mathcal{F}_1]$, (2) adding $T[\mathcal{I}_1]$, and (3) further incorporating $T[\mathcal{F}_2]$, the full version employed in \thiswork. This breakdown allows us to observe the effect of each added term on the accuracy of the approximation.

For all patterns, \fapx of \thiswork achieves orders of magnitude smaller KL divergence than \funi, indicating a much closer approximation to \fideal. We also find that all three terms are necessary. For instance, in 4-clique, using only the first term yields a KL divergence slightly larger than \funi. By adding the term $T[\mathcal{I}_1]$,  we decrease the value to nearly 0. Similarly, for 4-chain, including the final term $T[\mathcal{F}_2]$ further boosts the deduction in KL divergence.

\cref{fig:fapprox_performance_right} presents the number of samplers required for convergence with incremental addition of the terms. Here, we include larger patterns for which it is not feasible to compute \fideal. As shown, having all three terms yields clear benefits. However, as discussed in \cref{sec:method:approx}, the marginal improvement from adding the third term is the smallest. This observation aligns with the notion that incorporating higher-order terms leads to diminishing returns, since edges further along the pattern introduce greater uncertainty.


\subsection{Comparison with MCMC based method}
\label{sec:eval:mcmc}


\begin{figure}[t]
    \centering
    \includegraphics[width=0.49\textwidth]{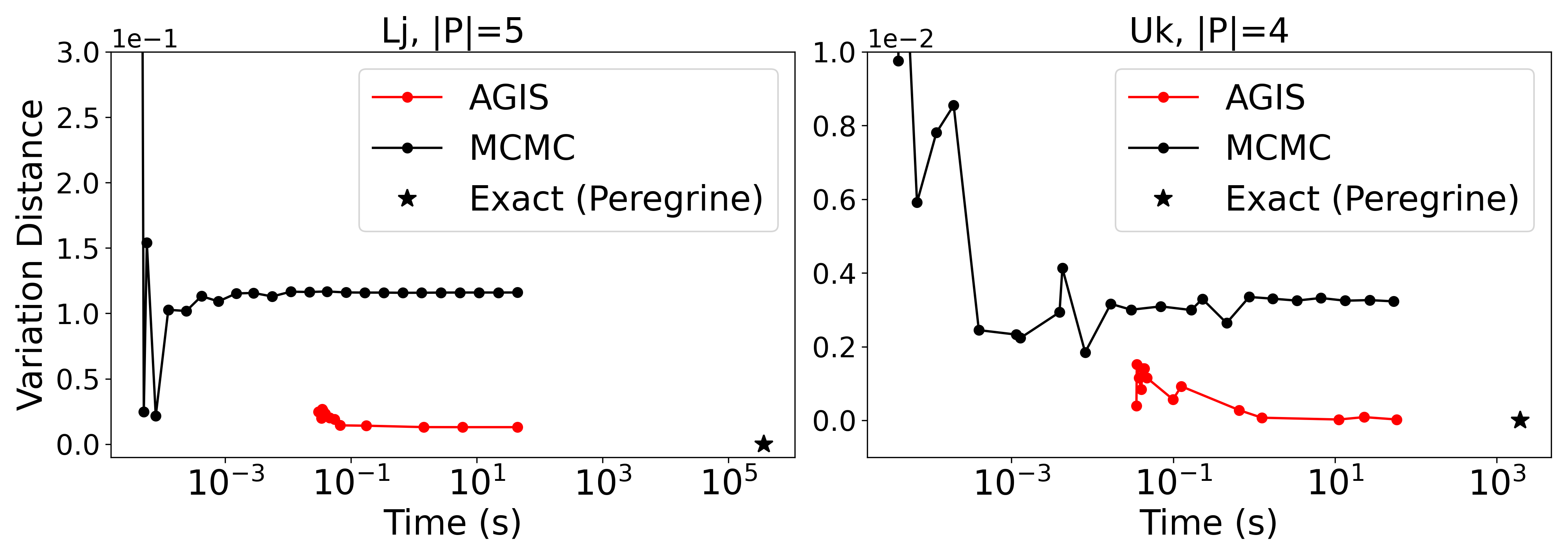} 
    \caption{Variation distance to the true concentration.}
    
    \label{fig:mcmc}
\end{figure}

Markov Chain Monte Carlo (MCMC)–based counting techniques~\cite{mc1,mc2,mc4,mc6} provide an alternative to \texttt{NS}-based sampling methods for estimating the concentration, the proportion of a specific pattern among all patterns of the same size in a given graph. 
These techniques construct a higher-order auxiliary graph whose vertices correspond to embeddings in the original graph.
Two vertices of the auxiliary graph are adjacent if their embeddings differ by a small local modification, such as replacing, deleting, or adding a vertex. 
An MCMC algorithm then performs a random walk on this auxiliary graph until mixing, using the visitation frequencies to approximate the concentration.

\cref{fig:mcmc} compares \thiswork with an MCMC-based counting algorithm on size-five patterns over the \texttt{Lj} graph and size-four patterns over the \texttt{Uk} graph. 
We implemented SRW2, the MCMC-based algorithm described in~\cite{mc6}, which returns the estimated concentration for a specified pattern size. 
For \thiswork, there are a total of 21 connected non‑isomorphic size‑five patterns and 6 connected non‑isomorphic size‑four patterns. We enumerated each pattern individually to obtain its approximate count, and then normalized these counts by the total number of patterns of the corresponding size to derive their concentrations.
Ground-truth concentrations were obtained with the exact system Peregrine. 
Accuracy was measured by variation distance, defined as one half of the \(L_{1}\) distance between concentration vectors. 
The figure shows that, while MCMC methods enjoy strong theoretical guarantees, \thiswork produces much more accurate estimates within the same time budget. 
Specifically, at 1\,s, the MCMC method exhibits variation distances that are 8.3\(\times\) and 41.7\(\times\) larger than those of \thiswork.


\subsection{Execution Time Breakdown}
\label{sec:eval:preprocess}

\begin{figure}[t]
    \centering
    \begin{subfigure}{0.49\linewidth}
        \centering
        \includegraphics[width=\linewidth]{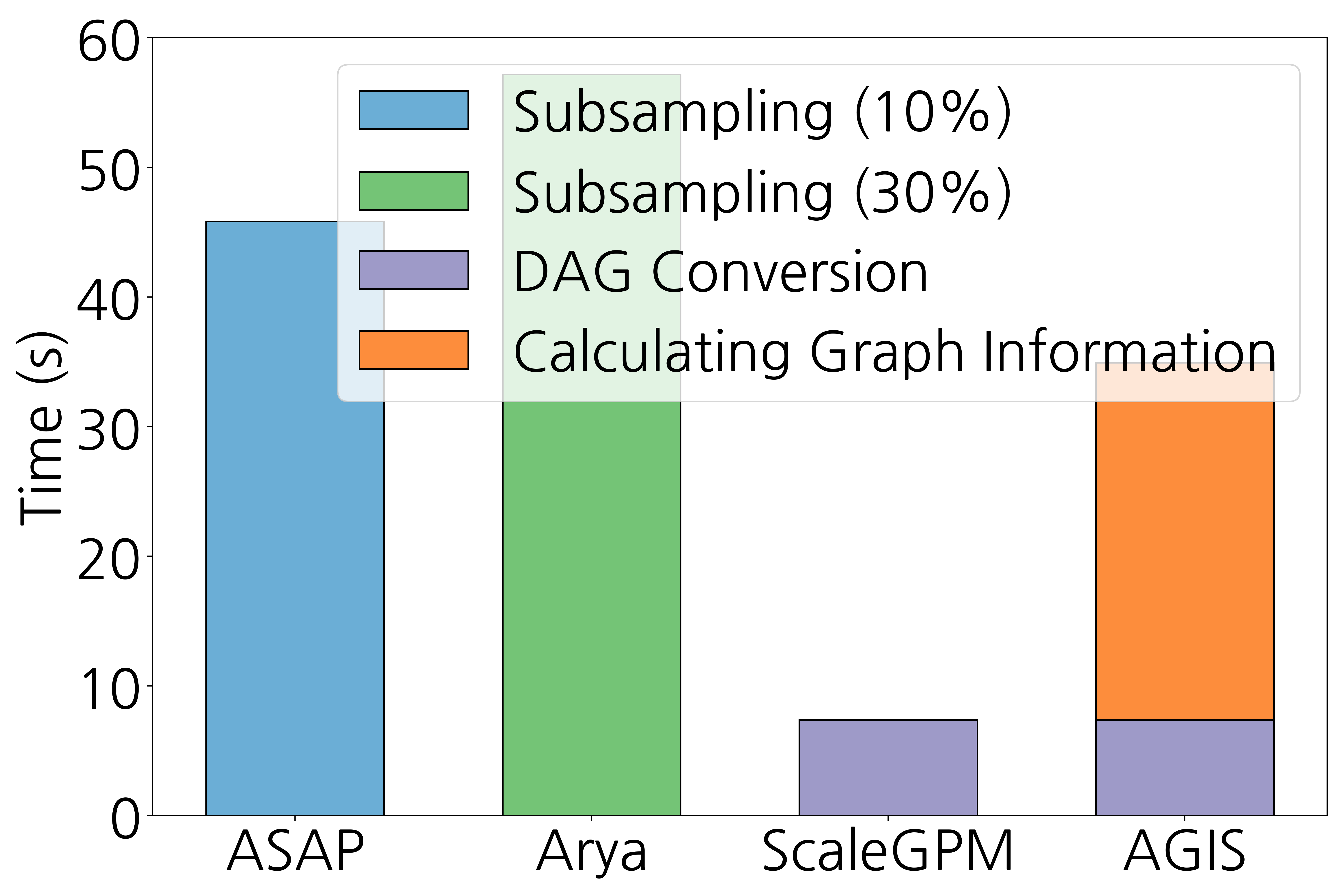}
        \caption{\texttt{Tw} preprocessing time. }
        \label{fig:time preprocessing}
    \end{subfigure}
    \begin{subfigure}{0.49\linewidth}
        \centering
        \includegraphics[width=\linewidth]{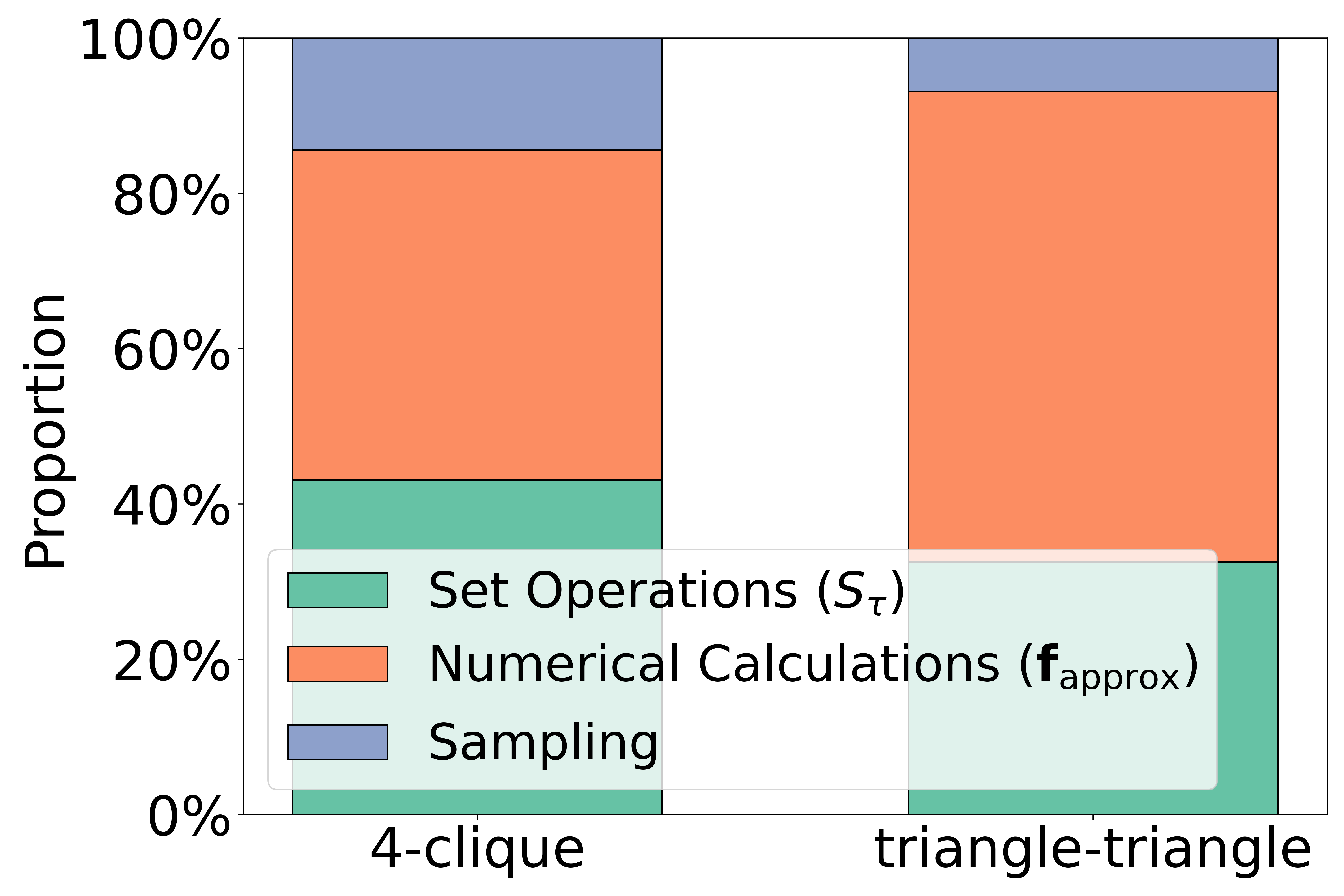}
        \caption{\texttt{Tw} mining time breakdown.}
        \label{fig:time ratio}
    \end{subfigure}
    \caption{Execution time analysis. }
    \label{fig:exectime}
\end{figure}

\begin{table}[t]
\centering
\footnotesize
\begin{tabular}{lcccc}
\toprule
\textbf{Graph} & \texttt{Lj} & \texttt{Uk} & \texttt{Tw} & \texttt{Fs} \\
\midrule
\textbf{Total Time (s)} & 0.477 & 3.069 & 34.93 & 55.37 \\
\bottomrule
\end{tabular}
\caption{
Preprocessing time across different graph datasets.}
\label{tab:total_preproc_time}
\end{table}

We now consider the time spent on preprocessing. For ASAP and Arya, a subsampling step must be performed once per given graph. This step involves randomly sampling edges at a specified ratio, converting them into an undirected graph representation, and then refining them into a CSR format. For ScaleGPM, a directed acyclic graph (DAG) must be constructed for orientation optimization. In the case of \thiswork, in addition to the DAG construction, we must also compute the required graph information (\cref{sec:method:preprocessing}).

\cref{fig:time preprocessing} presents the preprocessing times for the four AGPM systems. By employing the sampling technique described in \cref{sec:method:preprocessing}, the preprocessing phase for \thiswork is not prohibitively slow, and remains comparable to that of ELP-based methods. 
For completeness, \cref{tab:total_preproc_time} reports the total preprocessing time for each graph. 
This one‑time overhead is amortized over all pattern queries and thus negligible in practice, particularly when mining complex patterns or when mining under tight error bounds.

Lastly, we present a detailed timing breakdown of \thiswork's runtime in \cref{fig:time ratio}. We partition the execution into three components: (1) set operations for constructing the candidate set $\mathbf{S}_\tau$, (2) numerical calculations for \fapx and (3) sampling via building a cumulative distribution. The analysis indicates that the computation of \fapx accounts for slightly more than half of the total runtime. However, as demonstrated in \cref{sec:eval:overall} and \cref{sec:eval:fapprox}, this additional cost is justified, as a meticulously constructed \fapx leads to a substantial improvement in overall runtime.

\section{Related Work}
\label{sec:related_work}
\textbf{Exact Graph Pattern Mining Systems}. There have been numerous attempts to develop efficient exact Graph Pattern Mining systems. Systems such as~\cite{peregrine, graphzero, rstream, dwarvesgraph, sandslash}, use various optimization techniques to improve performance.
For instance, DwarvesGraph~\cite{dwarvesgraph} uses a decomposition algorithm and computes the counts for each decomposed pattern. GraphZero~\cite{graphzero} uses a relabeling technique to generalize the orientation algorithm to arbitrary patterns. Peregrine~\cite{peregrine} adopts pattern-aware algorithms to prune early, thus bypassing expensive isomorphism and canonicality checks. 
\cite{C4,C5} develops a combinatorial framework for efficient counting.
Notably, \cite{automine, graphpi} employ a performance model with similar approximation logic to \thiswork. They approximate the size of set intersections using a global constant (e.g., the total number of triangles), whereas \thiswork leverages local fine grained information.

%

\noindent\textbf{Other Approximate Counting Schemes.} 
Beyond \texttt{NS}-based methods and MCMC-based methods, several alternative approaches have been developed for approximate subgraph counting. 
One example is the Color Coding (CC) approach~\cite{cc1, cc2, cc3, cc4, cc5}. The core idea is to assign each vertex a random color and focus on colorful matches, where all vertices in the pattern have distinct colors. This counting step can be done efficiently via dynamic programming, and by repeating the color assignment and counting procedure multiple times, the approximation can be refined.

%
%
Another line of research focuses on sparsifying the graph before applying exact counting procedures. 
Various studies have introduced distinct sparsification techniques to approximate specific patterns, including triangles~\cite{BES, CS}, cliques~\cite{sparsification_for_clique_count}, 5-cycles~\cite{sparsification_for_5_cycle}, and butterfly patterns~\cite{sparsification_for_butterfly}.
In a similar vein, ScaleGPM~\cite{scalegpm} introduces a ``loose mode'' that leverages sparsification when its primary strategy is anticipated to fail or become excessively time-consuming.

Some streaming algorithms~\cite{C1,C2} maintain an edge reservoir by assigning each edge an inclusion probability proportional to its marginal contribution to the overall pattern count. This is analogous to \thiswork, in which optimal sampling probabilities are proportional to the number of pattern instances incident on each vertex.



\noindent\textbf{Counting Across Different Graph Types.}
While homogeneous graphs were considered in this paper, counting on heterogeneous graphs and hypergraphs is another important line of research. 
Type‑aware counting techniques have been developed to efficiently handle constraints on heterogeneous graphs. 
These include combinatorial methods that exploit algebraic relationships to avoid exhaustive enumeration~\cite{Ahmed2017,Gu2018,Rossi2019,Rossi2020a,Yu2024}, discriminative mining approaches that discover frequent typed patterns~\cite{Zong2015,Fang2021}, and sampling‑based estimators~\cite{Rossi2020b,Shin2024}.

For hypergraphs and simplicial complexes, recent systems~\cite{Yang2023,su2023} leverage hyperedge features and parallelism. 
A growing body of work proposes sketch‑based approximations for counting small sub‑hypergraphs at scale~\cite{Bressan2025,Kallaugher2018,Rodl2005,Montgomery2024}. 
Recent studies adapt random‑walk–based sampling~\cite{Kim2023,Kim2025,Beigy2024,Roth2020,Samorodnitsky2023}.

\section{Conclusion}

In this paper, we introduced \thiswork, a fast approximate graph pattern mining system that leverages a novel \sampling technique. By integrating structural properties from both data and pattern graphs, \thiswork constructs an approximate ideal sampling distribution and then adaptively decides when to apply it for optimal efficiency. Experiments on diverse datasets show that \thiswork outperforms state-of-the-art baselines by more than an order of magnitude (28.5$\times$ in geometric mean), achieving rapid convergence and scaling to graphs with tens of billions of edges.

\begin{acks}
This work was supported by 
Institute of Information \& communications Technology Planning \& Evaluation (IITP) 
(RS-2024-00395134, 
RS-2024-00347394, 
RS-2023-00256081, 
RS-2021II211343, 
RS-2025-00564840). 
Jinho Lee is the corresponding author.
\end{acks}

\balance
\bibliographystyle{ACM-Reference-Format}
\bibliography{references}

\end{document}